\documentclass[apj]{emulateapj}
\usepackage{apjfonts}
\usepackage[usenames,dvipsnames]{xcolor}
\definecolor{blue}{rgb}{0,0,1}
\usepackage[hyperfootnotes=false]{hyperref}
\hypersetup{
  colorlinks,
  citecolor=blue,
  linkcolor=blue,
  urlcolor=Green}

\shortauthors{Liu, Guillochon, Lin \& Ramirez-Ruiz}

\begin{document}
\title{On the Survivability and Metamorphism of Tidally Disrupted Giant Planets:
the Role of Dense Cores}

\submitted{Accepted by \apj \ \  2012 November 6}

\author{Shang-Fei Liu\altaffilmark{1}$^*$, James
  Guillochon\altaffilmark{2}, Douglas N. C. Lin\altaffilmark{1,2} and 
  Enrico Ramirez-Ruiz\altaffilmark{2}}
\altaffiltext{*}{E-mail: \href{mailto:liushangfei@pku.edu.cn}{liushangfei@pku.edu.cn}}
\altaffiltext{1}{Kavli Institute for Astronomy and Astrophysics and 
Department of Astronomy, Peking University, Beijing 100871, China}
\altaffiltext{2}{Department of Astronomy and Astrophysics, 
University of California, Santa Cruz, CA 95064, U.S.A.}

\begin{abstract}
A large population of planetary candidates in short-period
orbits have been found recently through transit searches, mostly
with the {\it Kepler} mission. Radial velocity surveys have also
revealed several Jupiter-mass planets with highly eccentric
orbits. Measurements of the Rossiter-McLaughlin effect indicate
that the orbital angular momentum vector of some planets is
inclined relative to the spin axis of their host stars. This
diversity could be induced by post-formation dynamical processes
such as planet-planet scattering, the Kozai effect, or secular
chaos which brings planets to the vicinity of their host stars.
In this work, we propose a novel mechanism to form close-in
super-Earths and Neptune-like planets through the tidal
disruption of gas giant planets as a consequence of these
dynamical processes. We model the core-envelope structure of gas
giant planets with composite polytropes which characterize the
distinct chemical composition of the core and envelope. Using
three-dimensional hydrodynamical simulations of close encounters
between Jupiter-like planets and their host stars, we find that
the presence of a core with a mass more than ten times that of
the Earth can significantly increase the fraction of envelope
which remains bound to it. After the encounter, planets with
cores are more likely to be retained by their host stars in
contrast with previous studies which suggested that coreless
planets are often ejected. As a substantial fraction of their
gaseous envelopes is preferentially lost while the dense
incompressible cores retain most of their original mass, the
resulting metallicity of the surviving planets is increased. Our
results suggest that some gas giant planets can be effectively
transformed into either super-Earths or Neptune-like planets
after multiple close stellar passages. Finally, we analyze the
orbits and structure of known planets and {\it Kepler}
candidates and find that our model is capable of producing some
of the shortest-period objects.
\end{abstract}

\keywords{hydrodynamics --- star-planet interaction --- gas giant planets: internal structure --- super-Earths --- planetary systems: formation, population}

\section{Introduction}
In contrast to the kinematic architecture of our solar system,
there is a population of recently discovered exoplanets or
planetary candidates that have orbital periods ranging from days
to weeks. Depending on their masses, these close-in planets are
commonly referred to as hot Jupiters or super Neptunes. Their
relative abundance in the period distribution comes as the
result of observational bias as the current radial velocity and
transit surveys are more well suited for their detection than
the identification of planets with longer period and lower
masses. Recently, the {\it Kepler} mission has extended the
detection limit down to sub-Earth size objects, and unveiled a
rich population of close-in super-Earth and sub-Neptune
candidates (defined in terms of their sizes) around solar type
stars \citep{Batalha:2012uq}.

The origin of these close-in planets remains poorly understood.
A widely adopted scenario is based on the assumption that all
gas giant planets formed beyond the snow line a few AU from
their host star \citep{Pollack:1996zr}, with the progenitors of
hot Jupiters undergoing substantial inward migration through
planet-disk interaction \citep[see, e.g.,][]{Lin:1996ey,
Ida:2004ko, Papaloizou:2006lr}. This mechanism naturally leads
to the formation of resonant gas giants \citep{Lee:2002ef} and
coplanarity between the planets' orbits and their natal disks.
However, measurements of the Rossiter-McLaughlin effect
\citep{Ohta:2005qy} reveal that the orbits of a sub population
of hot Jupiters (around relatively massive and hot main sequence
stars) appear to be misaligned with the spin of their host stars
\citep{Winn:2010dr, Schlaufman:2010kx}. As the stellar spin is
assumed to be aligned with that of their surrounding disks
\citep{Lai:2011vn}, the observed stellar spin-planetary orbit
obliquity poses a challenge to the disk-migration scenario for
the origin of hot Jupiters \citep{Triaud:2010hr, Winn:2011fa}.

In order to reconcile the theoretical predictions with the
observations, some dynamical processes have been proposed, such
as the Kozai mechanism \citep{Kozai:1962fk, Takeda:2005kx,
Matsumura:2010ys, Naoz:2011vn, Nagasawa:2011dp}, planet-planet
scattering \citep[][]{Rasio:1996kx, Chatterjee:2008gd,
Ford:2008vn} or secular chaos \citep{Wu:2011df}, all of which
operate after the gas is depleted and the onset of dynamical
instability can produce highly eccentric orbits and considerably
large orbital obliquity. The observed eccentricity distribution
of extra-solar planets with periods longer than a week and
masses larger than that of Saturn has a median value noticeably
deviated from zero. Presumably they obtained this eccentricity
through dynamical instability after the depletion of their natal
disks \citep{Lin:1997ig, Zhou:2007fk, Chatterjee:2008gd,
Juric:2008uq}, as the eccentricity damping would suppress such
an instability if they were embedded in a gaseous disk
environment.

Some of these processes can produce planets that lie on nearly
parabolic orbits. As their eccentricity approaches unity,
planets with a semimajor axis of a few AU undergo close
encounters with their host stars. At their pericenters, tides
raised by the host star dissipate orbital energy into the
planet's internal energy, resulting in the shrinkage of their
semimajor axes. The repeated subsequent encounters may lead to
the circularization of their orbits \citep{Press:1977ys}, and
provided there is no mass loss, the planet's long-term orbital
evolution may be modeled analytically \citep{Ivanov:2007kx}.
However, when giant planets approach their host stars within
several stellar radii, the tidal force may become sufficiently
intense that it can lead to mass loss or tidal distruption. One
particular example is WASP-12b \citep{Li:2010lr}, which is being
tidally distorted and is continuously losing its mass.

Hydrodynamical simulations have been carried out by
\citet[][hereafter FRW]{Faber:2005lr} and \citet[][hereafter
GRL]{Guillochon:2011rt} to study the survivability and
orbital evolution of a Jupiter-mass planet disrupted by a
Sun-like star. In the description of the relative strength of
the tidal field exerted on a planet by the host star, it is
useful to define a characteristic tidal radius as 
\begin{equation}
r_{\rm t} \equiv \left(\frac{M_*}{M_{\rm P}}\right)^{1/3} R_{\rm P}, 
\label{equ:tidalradius}
\end{equation}
where $M_{\rm P}$ and $R_{\rm P}$ are the planetary mass and
radius, and $M_*$ is the stellar mass (not to be confused with
the Hill radius and Roche radius, which in this context commonly
refer to a separation distance as measured from the the center
of mass of the secondary). At this separation, the
volume-averaged stellar density equals to the planetary mean
density, i.e. $r_{\rm t} \simeq 1 \ R_{\odot}$ in this case. Our
previous simulations of single nearly parabolic (with $e \simeq
1$) encounters show the existence of a mass-shedding region
demarcated by $r_{\rm p}/r_{\rm t} \lesssim 2$, where $r_{\rm
p}$ is periastron distance. The planet's specific orbital
binding energy after the (either parabolic or highly elliptical)
encounters is smaller for larger impact parameters
($\beta=r_{\rm t}/r_{\rm p}$), despite an enhanced stellar tidal
perturbation. Within a sufficiently close range, planets are
ejected due to mass and energy loss near periastron.

For the more distant periastron encounters, we also investigated
planet's response after multiple passages (see GRL, Section
3.2). We considered orbits with $e=0.9$ and $r_{\rm p} / r_{\rm
t} \gtrsim 2$ and showed that successive encounters can enhance
planetary mass and energy changes. We found a critical
periastron separation $r_{\rm p} = 2.7 \ r_{\rm t}$ within which
no planet can avoid destruction. However, this critical value
only places a lower limit on non-destructive tidal interactions,
as the accumulation of energy required to destroy a planet at
wider separations occurs over a much longer time scale, which
has not yet been investigated. We also noted that the semimajor
axes of several known exoplanets are less than twice this
critical separation. If they were scattered to the proximity of
the star on a highly eccentric orbit ($e \gtrsim 0.9$), the
initial periastron separation would be less than $2.7 \  r_{\rm
t}$, and thus they would have already been destroyed. We
suggested that either these planets were scattered from a
distance that is substantially closer to the host star than the
snow line, or they were scattered to a further separation and
then later migrated inward under the influence of tidal
interaction with their host stars to their present positions.

To summarize, the observed inner edge of hot Jupiters seems to
suggest they were tidally circularized \citep{Ford:2006fk,
Hellier:2012uq}, as the hydrodynamical simulations (FRW and GRL)
showed that tidal dissipation within the planet alone either
results in the planet's ejection or disruption. In this work, we
re-examine the disruption and retention of gas giant planets
during their close encounters with their host stars by taking
into account the presence of their dense cores. This possibility
is not only consistent with the internal structure of Saturn
\citep[and to a much less certain extent in
Jupiter,][]{Guillot:2004hc}, but is also consistent with the
widely adopted core accretion scenario \citep{Pollack:1996zr}.
We show that presence of a core with mass as small as
$10 \ M_{\earth}$, e.g. 3\% of a Jupiter-like planet's total mass,
the planet has a far greater chance of survival, even with a
mass loss comparable to the mass within its own envelope. We
also consider the possibility that the tidal disruption
mechanism may be an efficient way to transform a Jupiter-mass
planet into a close-in super-Earth or Neptune-like object, which
potentially may explain the existence of some of the inner edge
of close-in planets.

Our paper is organized as follows. In
Section~\ref{sec:compositepoly}, we introduce a composite
polytrope model for planets with cores. Our setup for
hydrodynamic simulations is described in
Section~\ref{sec:methods}. We present our simulation results in
Section~\ref{sec:results}. In Section~\ref{sec:dis}, we first
discuss the adiabatic responses of mass-losing composite
polytropes and explain the enhanced survivability of planets
with cores as suggested by our numerical results, and search for
potential candidates of tidally disrupted planets in the current
exoplanet sample (including {\it Kepler} candidates). We
summarize our work and probe the future directions in
Section~\ref{sec:summary}.

\section{A Composite Polytrope Model for Gas Giant Planets with Cores}
\label{sec:compositepoly}
\begin{figure*}
  \centering
  \includegraphics[width=\linewidth,clip=true]{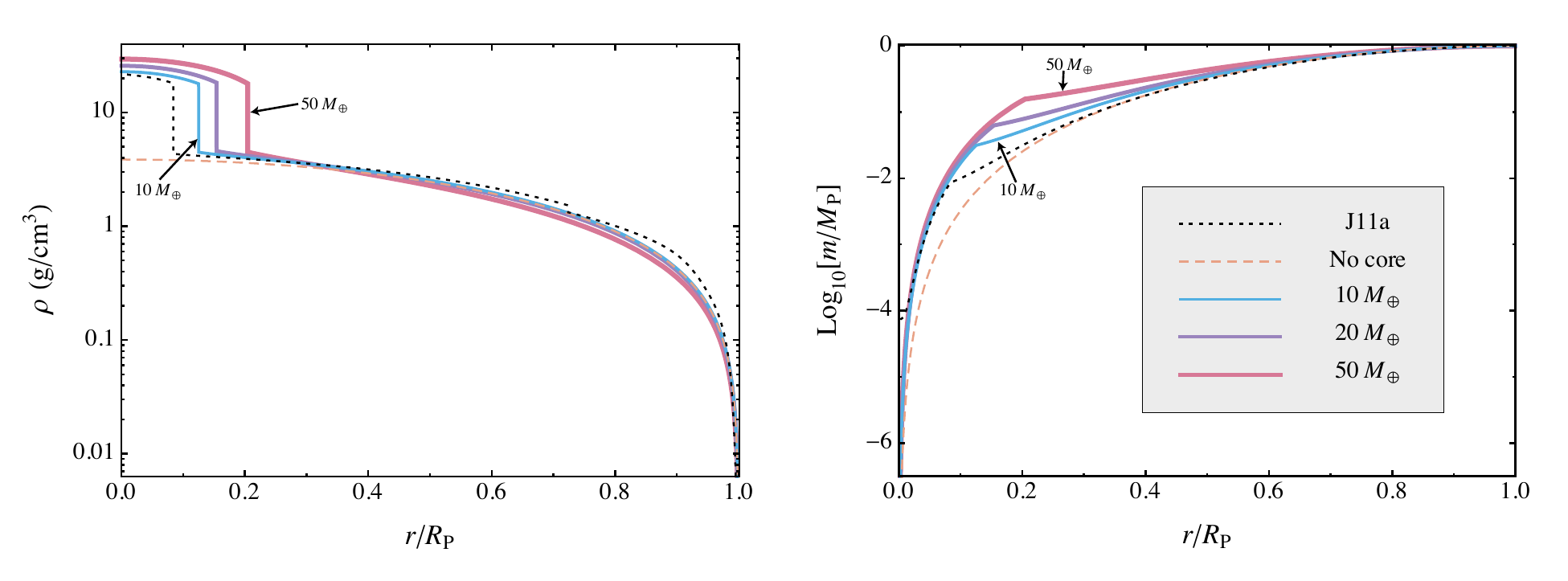}
\caption{One-dimensional profiles of density (left) and enclosed
mass (right) of a composite polytropic model ($n_1=0.5$ and
$n_2=1$) for a Jupiter-mass planet with no core, a $10
M_{\oplus}$ core, a $20 M_{\oplus}$ core, and a $50 M_{\oplus}$
core. Note the density jump (left panel) and the discontinuity
of the derivative of mass distribution (right panel) at the
core-envelope interfaces are a result of the difference in
molecular weight between the two zones, $\mu_1=4\mu_2$. The
single $n = 1$ polytropic model (dashed line) used in our
previous simulations together with a profile of Jupiter with a
2.75 $M_{\oplus}$ core (dotted line) taken from
\citet{Nettelmann:2008fk} are plotted for comparison.}
\label{fig:1dprofile}
\end{figure*}

The core-envelope structure of gas giant planets is determined
by the equation of state (EOS), their metal content, and their
thermal evolution \citep{Guillot:2004hc}. For computational
simplicity, we approximate this structure by a composite
polytrope model. This class of models is thoroughly described in
\citet[][]{Horedt:2004uq}. Previously, a set of composite
$n_1=3$ and $n_2=1.5$ polytropes has been used to represent the
radiative core and the convective envelope of stars
\citep{Rappaport:1983fj}. By finding the intersection of the
solutions of the Lane-Emden equation in the core and envelope on
the U-V plane \citep[see e.g. chapter 21
of][]{Kippenhahn:1994fk}, the overall physical properties
envelope in stars can be calculated. In this paper, we adopt
this approach with the incorporation of different species to
model the transition in the composition and EOS at the
core-envelope interface in giant planets (see
Figure~\ref{fig:1dprofile}).

The polytropic approximation is simple to use because the pressure $P$ 
is a power-law function of the density $\rho$ only
\begin{equation}
  P=K \rho^\gamma=K\rho^{(n+1)/n},
\label{poly}
\end{equation}
where $K$ is a constant. We denote the quantities related to the
core and envelope by subscripts 1 and 2, respectively. To model
the composite polytropic planet, we choose the polytropic
indices to be $n_1=0.5$ and $n_2=1$ in the core and envelope,
corresponding to $\gamma_1=3$ and $\gamma_2=2$.

Following \citet{Rappaport:1983fj}, we express the densities and pressures as 
\begin{equation}
\rho_1=\rho_{\rm 1c}\theta_1^{n_1},\; P_1=K_1\rho_{\rm 1c}^{(n_1+1)/n_1}\theta_1^{n_1+1},
\end{equation}
\begin{equation}
 \rho_2=\rho_{\rm 2i}\theta_2^{n_2},\; P_2=K_2\rho_{\rm 2i}^{(n_2+1)/n_2}\theta_2^{n_2+1}.
\end{equation}
The subscripts $\rm c$ and $\rm i$ denote quantities evaluated
at the planetary center and core-envelope interface,
respectively, and $\theta$ is a dimensionless variable which
satisfies the Lane-Emden equation
\begin{equation}
\frac{1}{\xi^2}\frac{d}{d\xi}\left(\xi^2\frac{d\theta}{d\xi}\right)=-\theta^n.
  \label{lane_emden}
  \end{equation}
The dimensionless length $\xi$ is defined by $\xi=r/a$, where
\begin{equation}
  a_1=\left[\frac{(n_1+1)K_1}{4\pi G}\right]^{1/2}\rho_{\rm 1c}^{-\frac{(n_1-1)}{2n_1}} , 
\end{equation}
\begin{equation}
  a_2=\left[\frac{(n_2+1)K_2}{4 \pi G}\right]^{1/2}\rho_{\rm 2i}^{-\frac{(n_2-1)}{2n_2}}.
\end{equation}
We can obtain the mass contained within radius $r=a\xi$ by
\begin{equation}
  m_1\left(\xi_1\right)=-4\pi\rho_{\rm 1c} a_1^3\left(\xi_1^2\theta_1'\right) , 
\end{equation}
\begin{equation}  
  m_2\left(\xi_2\right)=-4\pi\rho_{\rm 2i} a_2^3\left(\xi_i^2\theta_2'\right),
\label{mass}
\end{equation}
where we use the notation $\theta '$ to denote the derivative
$d\theta/d\xi$. The continuity of density, pressure, radius and
mass at the interface yields
\begin{equation}
\label{continuity1}
  \frac{\xi_{\rm 1i}\theta_{\rm 1i}^{n_1}}{\theta_{\rm 1i}'\mu_1} = \frac{\xi_{\rm 2i}\theta_{\rm 2i}^{n_2}}{\theta_{\rm 2i}'\mu_2},
\end{equation}
\begin{equation}
\label{continuity2}
  \left(\frac{n_1+1}{n_2+1}\right) \frac{\xi_{\rm 1i}\theta_{\rm 1i}'}{\theta_{\rm 1i}\mu_1} = \frac{\xi_{\rm 2i}\theta_{\rm 2i}'}{\theta_{\rm 2i}\mu_2},
\end{equation}
where $\mu_1$ and $\mu_2$ are the mean molecular weight in the
core and the envelope.

The Lane-Emden equation with $n=0.5$ in the core can be
integrated from the center of the planet outward directly, with
the inner boundary conditions
\begin{equation}
  \label{innerboundary}
  \theta_1\left(0\right)=1 \;\; \textrm{and} \;\; \theta_1'\left(0\right)=0,
\end{equation}
which imply that the central density is finite and its
derivative vanishes \citep[chapter 19 of][]{Kippenhahn:1994fk}.
However, to determine the solution of the Lane-Emden equation in
the envelope, we need to specify a cut-off $\xi_{\rm 1i}$ of the
solution $\theta_1\left(\xi_1\right)$ in the core, so
$\theta_{\rm 1i}$ and $\theta_{\rm 1i}'$ can be calculated in a
straightforward manner. Consequently, $ \xi_{\rm 2i}$ and
$\theta_{\rm 2i}$ can be evaluated using the continuity
equations (\ref{continuity1}) and (\ref{continuity2}).

\begin{equation}
  \label{cutoff1}
  \xi_{\rm 2i}=\left(\frac{n_1+1}{n_2+1} \; \frac{\theta_{\rm 1i}^{n_1-1}}{\theta_{\rm 2i}^{n_2-1}}\right)^{1/2} \frac{\mu_2}{\mu_1} \xi_{\rm 1i}
\end{equation}

\begin{equation}
  \label{cutoff2}
  \theta_{\rm 2i}'=\left(\frac{n_1+1}{n_2+1} \right) \;\frac{ \xi_{\rm 1i} \theta_{\rm 1i}' }{\xi_{\rm 2i}} \; \frac{\theta_{\rm 2i} }{\theta_{\rm 1i}} \;\frac{\mu_2}{\mu_1}
\end{equation}

For simplicity we take $\theta_{\rm 2i}=1$, and then equations (\ref{cutoff1}) and (\ref{cutoff2}) become 
\begin{equation}
  \label{cutoff3}
  \xi_{\rm 2i}=\left(\frac{n_1+1}{n_2+1} \; \theta_{\rm 1i}^{n_1-1}\right)^{1/2} \frac{\mu_2}{\mu_1} \xi_{\rm 1i}
\end{equation}

\begin{equation}
  \label{cutoff4}
  \theta_{\rm 2i}'=\left(\frac{n_1+1}{n_2+1} \right) \;\frac{ \xi_{\rm 1i} \theta_{\rm 1i}' }{\xi_{\rm 2i} \theta_{\rm 1i}} \;\frac{\mu_2}{\mu_1}
\end{equation}
In this case, the solution $\theta_2\left(\xi_2\right)$ of Lane-Emden equation in the envelope is not finite at the origin, which poses no problem as it is not evaluated below $\xi_{\rm 2i}$.

\begin{deluxetable}{cccccc}
    \tablecolumns{6}
    \tablecaption{Parameters of composite polytrope models}
    \tablehead{\colhead{$M_{\rm core}$} & \colhead{$\xi_{\rm 1i}$} & \colhead{$\xi_{\rm 2i}$} & \colhead{$\rho_{\rm 1c}$ \tablenotemark{a}} & \colhead{$\rho_{\rm 2i}$ \tablenotemark{b}} & \colhead{$Q$\tablenotemark{c}}  \\ 
    \colhead{($M_{\oplus}$)} & & & \colhead{(${\rm g}/{\rm cm^3}$)} & \colhead{(${\rm g}/{\rm cm^3}$)}  } 
    \startdata
        10 & 1.571 & 0.3841  & 22.79 & 4.47 & 0.1251\\
        20 & 1.799 & 0.4619 & 25.69 & 4.57 & 0.1541 \\
        50 & 2.064 & 0.5734 & 29.43 & 4.47 & 0.2047
    \enddata
    \tablenotetext{a}{$\rho_{\rm 1c}$ is the central density of the model.}
    \tablenotetext{b}{$\rho_{\rm 2i}$ is the density of the envelope at the core-envelope interface.}
    \tablenotetext{c}{$Q=R_{\rm core}/R_{\rm P}$, where $R_{\rm core}$ and $R_{\rm P}$ are the core and planet radii, respectively.}
    \label{tab:compoly}
\end{deluxetable}

In this work, we generate three composite polytrope models for a
Jupiter-like planet with core masses of 10 $M_{\oplus}$,
20 $M_{\oplus}$ and 50 $M_{\oplus}$. The parameters of each model
are summarized in Table~\ref{tab:compoly}, where a constant
$\mu_1=4 \ \mu_2$ has been assumed. Figure~\ref{fig:1dprofile}
shows the density and mass distribution of these
 models (solid colored lines). The orange dashed line
indicates the single-layered polytrope model, and the black
dotted line shows a three-layer model for Jupiter taken from
\citet{Nettelmann:2008fk}, which includes a 2.75 $M_{\oplus}$
core. Though the models presented here have more massive cores,
our composite polytrope models generally fit the three-layer
model very well, whereas the single-layered polytrope model
fails to represent the high density of the core.

\section{Hydrodynamical Simulations of Tidal Disruption}
\subsection{Methods}\label{sec:methods}
\begin{figure*}
  \centering\includegraphics[width=0.75\linewidth,clip=true]{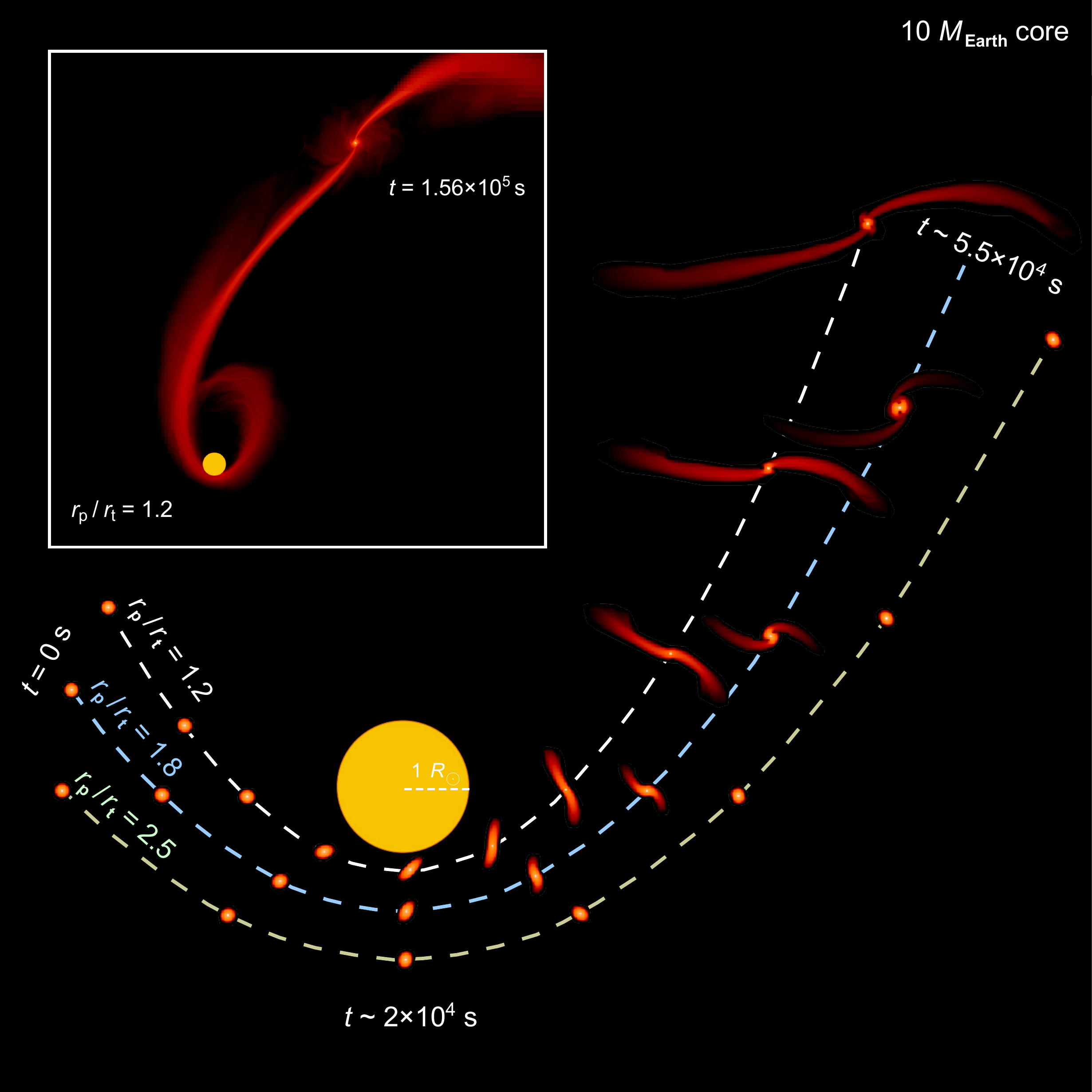}
\caption{Snapshots from several simulations of the tidal
disruption of a Jupiter mass planet with a $10 \ M_\earth$ core at
different periastron distances. The main panel superimposes the
trajectories (dashed lines) and the hydrodynamical evolution of
the planet in the stellar tidal field as it flies by the star
(from left to right), with the pericenter separations $r_{\rm
p}$ being is $1.2 \  r_{\rm t}$, $1.8 \ r_{\rm t}$ and $2.5 \ r_{\rm t}$,
respectively. After the encounter, mass is stripped from the
planet through the inner ($L_1$) and outer Lagrange point
($L_2$) and forms the two tidal streams. The material flowing
through $L_{1}$ then falls back to the host star and is
eventually accreted (inset panel). The yellow filled circles
represent the position and size of the star, taken here to be
equal to that of the Sun.}
  \label{fig:snapshot}
\end{figure*}

\begin{figure*}
  \centering
  \includegraphics[width=\linewidth,clip=true]{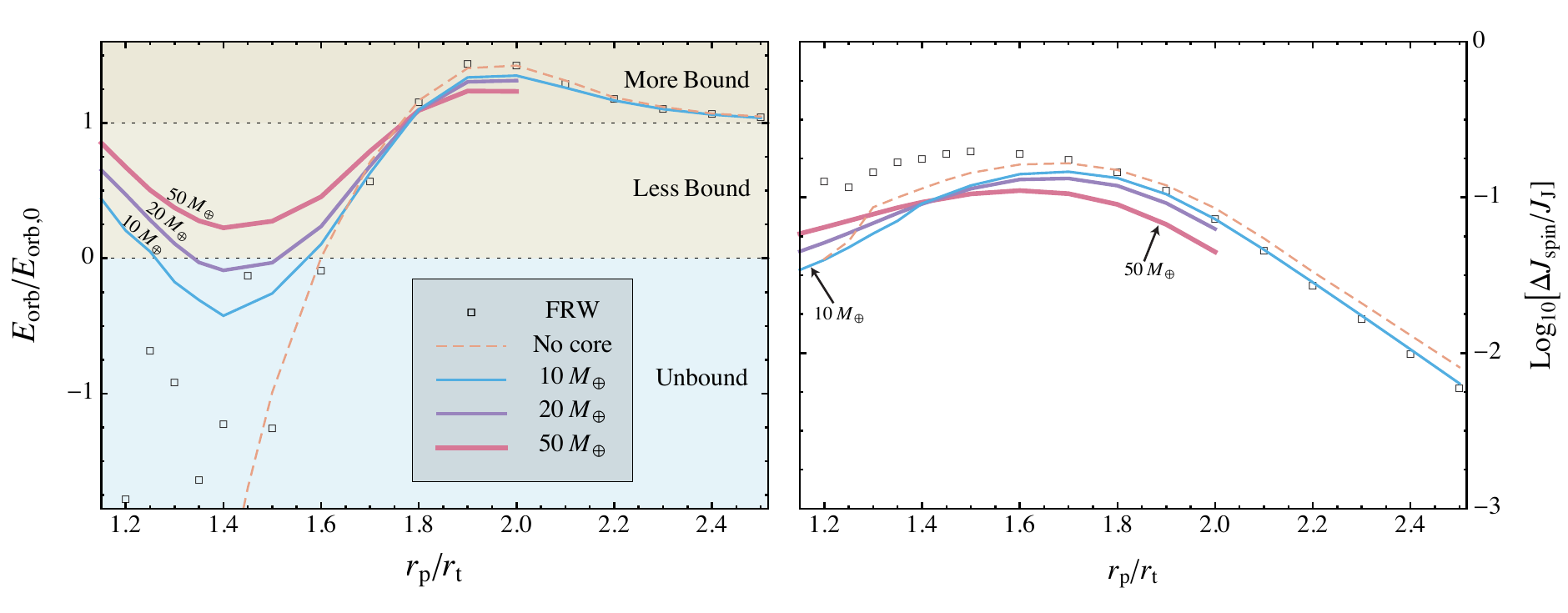}
\caption{Final specific orbital energy $E_{\rm orb}$ scaled to
the initial specific orbital energy $E_{\rm orb,0}$ (left panel)
and spin angular momentum $J_{\rm spin}$ scaled by the
characteristic angular momentum of Jupiter $J_{\rm J}^2=GM_{\rm
J}^3R_{\rm J}$ (right panel) as functions of periastron distance
$r_{\rm p}$ in units of tidal radius $r_{\rm t}$ after a single
near-parabolic encounter between a $M_{\rm P} = M_{\rm J}$
planet and a $M_* = 10^3 \  M_{\rm J}$ star. Open squares show the
data for coreless planets as presented in FRW, whereas the
dashed line shows the results of GRL. The three colored solid
lines show the results from disruption simulations of
Jupiter-like planets with core masses of 10, 20, and 50
$M_{\oplus}$, respectively. The filled regions in the left-hand
panel show the three possible outcomes: The planet either
becomes more bound, less bound, or completely unbound from its
parent star after the encounter.}
  \label{fig:EorbJspin}
\end{figure*}

We carry out numerical simulations to follow the hydrodynamic
response of gas giant planets during their close encounters with
their host stars. Our simulations are constructed based on the
framework of FLASH \citep{Fryxell:2000kx}, an adaptive-mesh,
grid-based hydrodynamics code \citep[a good introduction to
grid-based numerical methods is given in][]{Bodenheimer:2007kx}.
The simulation of tidal disruptions within the FLASH framework
was initially outlined in \citet{Guillochon:2009vn}. In that
work, the disruption of stars by supermassive black holes
(SMBHs) were simulated for the purpose of characterizing the
shock breakout signature resulting from the extreme compression
associated with particularly strong encounters. In GRL the code
was adapted to simulate the effects of strong tides on giant,
coreless planets after both single and multiple close-in
passages. Recently, \citep{Guillochon:2012uq, MacLeod:2012fk}
used this same code formalism to determine the feeding rate of
supermassive black holes from the disruptions of both
main-sequence and evolved stars at various pericenter distances.

In this work, we further extend the numerical framework
presented in the above references to include the ability to
simulate multi-layered objects, with each layer obeying a
separate equation of state (EOS). As before, we treat the star
as a point-mass \citep{Matsumura:2008yq}, and the simulations
are performed in the rest-frame of the planet to avoid issues
relating to the non-Galilean invariance (GI) of the Riemann
problem \citep{Springel:2010vn}. Our planets are modeled using
composite polytropes (as described in
Section~\ref{sec:compositepoly}), and we further assume that the
adiabiatic indices are equal to the polytropic indices. This
provides a reasonable approximation to the structures of
Jupiter-like planets \citep{Hubbard:1984ly}.

The total volume of the simulation box is $10^{13}\times
10^{13}\times 10^{13} \; \textrm{cm}^3$. The initial conditions
are identical to that of FRW and GRL to facilitate comparisons.
The planet is assumed to have a mass $M_{\rm P} = M_{\rm J}$ and
a radius $R_{\rm P}=R_{\rm J}$, where $M_{\rm J}$ and $R_{\rm
J}$ are Jovian mass and radius, respectively. The planets are
disrupted by a star with $M_* = 10^3 \ M_{\rm J} \simeq0.95 \ M_\sun$.
Thus, the tidal radius of the planet is $r_{\rm t}=10 \ R_{\rm
J} \simeq 0.995 \ R_\sun=0.00463$ AU.
 
We set the initial orbit of the incoming planet to have an
apastron separation $r_{\rm a}=10^4 \ R_{\rm J}$. At the onset of the
simulation, we set the distance of the planet from the star to
be 5 $r_{\rm t}$ such that tides are initially unimportant, and also
assume that the planet has no initial spin in the inertial
frame.

\begin{figure*}
  \centering
  \includegraphics[width=\linewidth,clip=true]{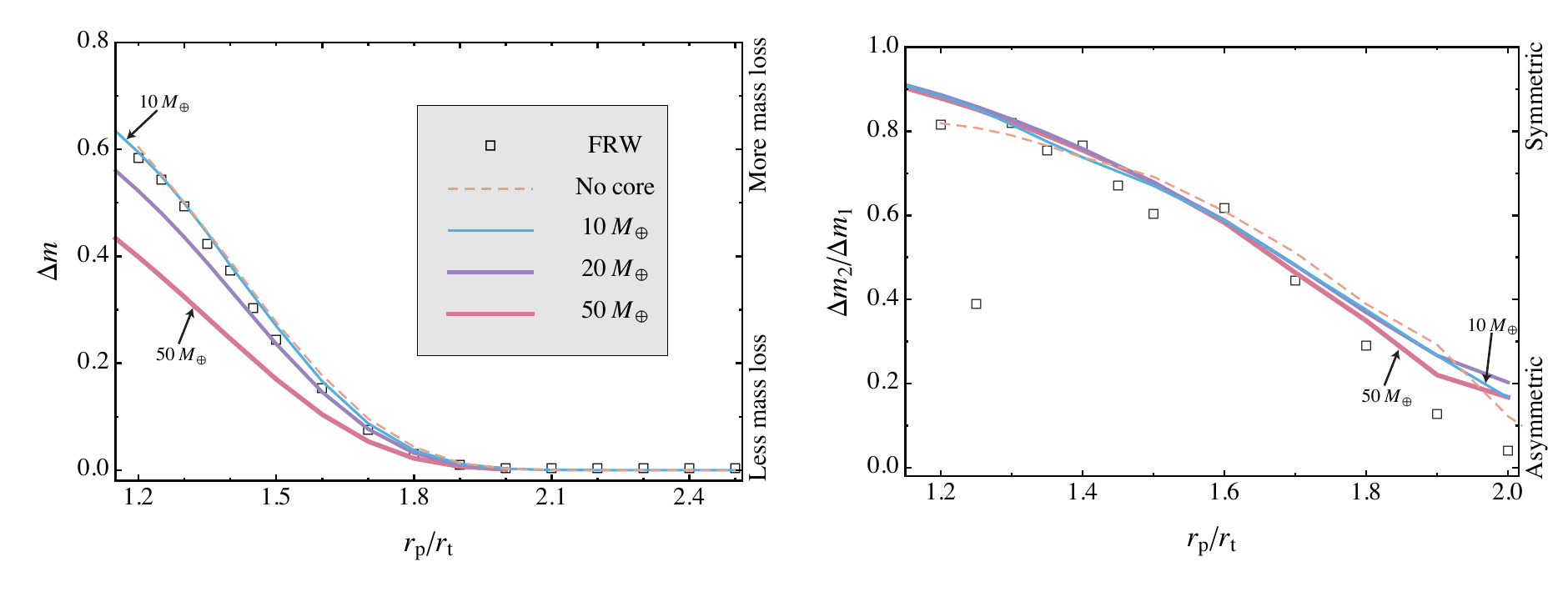}
\caption{Total mass fraction lost from Jupiter-like planets of
varying core masses as a function of $r_{\rm p} / r_{\rm t}$. The left panel
shows $\Delta m$ the total mass fraction unbound from the planet, whereas the
right panel compares the ratio between $\Delta m_2$, the fraction of mass 
lost from $L_{2}$, to $\Delta m_1$, the fraction of mass
 lost from $L_{1}$. The color scheme and line style are the same
as in Figure~\ref{fig:EorbJspin}.}
\label{fig:massloss}
\end{figure*}

\subsection{Results}\label{sec:results}
In total, we simulated 41 models with the three different core
masses listed in Table \ref{tab:compoly} and the initial
periastron distance $r_{\rm p}$ ranging from 1.15 to 2.5 $r_{\rm
t}$. A selection of simulations for $M_{\rm core} = 10 \
M_{\oplus}$ is illustrated in Figure~\ref{fig:snapshot}. To
explore the effect of the polytropic index of the core on the
dynamics of the encounter, we simulate one additional 10
$M_{\oplus}$ model using $n_1=0.01$, with $r_{\rm p} = 1.2 \
r_{\rm t}$. Despite the radically different adiabatic index, we
found less than a 1\% difference between the small $n_{1}$ and
our fiducial larger $n_{1}$ in terms of changes in orbital
energy and mass loss. Thus, our results are not sensitive to
relatively large value of $n_1$ used in our simulations, which
was chosen for numerical convenience. This approximation is not
expected to affect any of the results presented here, and should
remain appropriate as long as the core is much denser than the
envelope and can retain a significant amount of mass, which is
always true in our single passage simulations. However, this may
not be valid if the mass loss is large, as may be the case for
multiple passages. The reader is refer to
Section~\ref{sec:adiabatic} for a detailed explanation of the
adiabatic response of composite polytropes to mass-loss and its
relevance in describing the outcome of multiple passage
encounters.
  
\subsubsection{Final Orbits of Disrupted Giant Planets}
In all our simulations, the planet is placed on a bound orbit
with a negative orbital energy per unit mass $E_{\rm orb, 0}$.
We plot in the left panel of Figure~\ref{fig:EorbJspin} the
ratio of $E_{\rm orb} / E_{\rm orb, 0}$, where $E_{\rm orb}$ is
the energy per unit mass at the end of the simulation,
approximately 50 dynamical timescales after pericenter. A
planet's orbit is more (less) gravitationally bound to its host
star if this ratio attains a positive value greater (lesser)
than unity. A planet becomes unbound if this ratio attains a
negative value. For comparison with previous simulations, we
show the results obtained by FRW with open squares and those of
single-layered polytropes obtained by GRL with orange
dashed lines in Figure~\ref{fig:EorbJspin}. The results of the
new simulations with $10 M_\oplus$, $20 M_\oplus$ and $50
M_\earth$ cores are shown as colored solid lines.

We find that while the addition of a core produces qualitatively
different results than coreless models, there are no qualitative
differences when the core mass is varied for the values
investigated here. The results in the left panel of
Figure~\ref{fig:EorbJspin} show that for planets with a 10
$M_\oplus$ core, the magnitude of $E_{\rm orb}/ E_{\rm orb, 0}$
is greater than unity, i.e. the planet becomes more bound, for
all encounters with $r_{\rm p} / r_{\rm t} \gtrsim 1.75$,
approaching unity for distant encounters where $r_{\rm p} \gg
r_{\rm t}$. For encounters with $1.57 \lesssim r_{\rm p} /
r_{\rm t} \lesssim 1.75$, this ratio remains positive but below
unity, and thus these planets become less bound to their host
star. For $1.27 \lesssim r_{\rm p} / r_{\rm t} \lesssim 1.57$,
planets become unbound, whereas those with $1.15 \lesssim r_{\rm
p} / r_{\rm t} \lesssim 1.27$ lose approximately half of their
initial mass, yet remain bound to their host stars.

The non-monotonic relationship between $r_{\rm p}$ and the
change in orbital energy is considerably more complex than the
results presented in FRW or GRL, where a coreless giant planet
was assumed. Although the results of encounters with $r_{\rm
p}/r_{\rm t} \gtrsim 1.75$ are in general agreement with the
coreless models, the discrepancy is apparent for closer
encounters. These previous studies predicted that the planet
becomes successively less bound when the periastron separation
decreases, and all encounters with $ r_{\rm p} / r_{\rm t}
\lesssim 1.62$ lead to ejection. In contrast, our work suggests
that for orbits with $1.4 \lesssim r_{\rm p} / r_{\rm t}
\lesssim 1.7 $ the planet becomes successively less bound until
reaching a transitional point at $r_{\rm p} / r_{\rm t} \sim
1.4$. Interior to this separation, the trend is reversed, with
the orbit becoming less unbound until $r_{\rm p} / r_{\rm t}
\sim 1.27$, where the planet's orbital binding energy is
comparable to its initial binding energy. A similar but more
pronounced trend is found for planets with $20 M_\earth$ and $50
M_\earth$ cores. The more massive the core is, the more unlikely
the planet will be ejected.

For planets with a $50 \ M_\earth$ core, we find that a
Jupiter-mass planet cannot be ejected in all cases we
investigated with the assumed initial apastron, which we
presumed to be equal to the host star's ice line. The location
of the $E_{\rm orb} = 0$ crossing points as a function of
$r_{\rm p}$ (Figure~\ref{fig:EorbJspin}) depends on how the
planet's self-binding energy compares to its initial orbital
energy. Changing the size of the planet, the ratio of the mass
between the star and the planet, or the initial eccentricity can
alter the normalization of $E_{\rm orb}/E_{\rm orb,0}$. For
example, a smaller initial binding energy can facilitate more
planetary ejections, whereas an initially more bound planet may
not be capable of being ejected for any $r_{\rm p}$.

An intriguing aspect of the work presented here is that if a
dense core is present, a giant planet can remain bound to the
star within certain limits of periastron separation, whereas
previous simulations (e.g., FRW and GRL) suggested that planets
without a core are always ejected or destroyed if any mass is
lost during the initial inspiral. The presence of the core
permits planets to plunge deeply into their parent star's tidal
field and potentially survive as a close-in planet on a circular
orbit.

It is desirable to study how does the orbit of the tidally
disrupted planet evolve during subsequent encounters. But due to
the extremely long orbital period of the highly eccentric giant
planet ($T_{\rm orbit} / T_{\rm simulation} \sim 10^4$),
numerical simulations that try to directly follow several orbits
of the disrupted remnants are currently prohibitive. It is not
clear yet whether these (marginally) bound planets will be
circularized or ejected after several encounters. GRL simulated
the multiple passages with a lower eccentricity ($e = 0.9$), and
these planets were found to be destroyed eventually after
several close encounters. We do not repeat the multiple passage
simulation with a lower eccentricity in this work, however, in
Section~\ref{sec:adiabatic} we study the adiabatic response of
composite polytropes to mass-loss and use the results in
Section~\ref{sec:densecore} to explain why the presence of a
dense core helps prevent planets from being destroyed in
subsequent passages. This in stark contrast to the results of
our previous simulations which show that core-less planets are
always tidally destroyed.

The right panel of Figure~\ref{fig:EorbJspin} shows the change
in spin angular momentum of the planet $J_{\rm spin}$ scaled by
that of Jupiter. The planet is set to be non-rotating initially.
Our hydrodynamic simulations show for Jupiter-mass planets with
10 $M_\earth$ and 20 $M_\earth$ cores $J_{\rm spin}$ peaks at
around $1.7 \ r_{\rm t}$, and for those with 50 $M_\earth$ cores
$J_{\rm spin}$ peaks at around $1.6 \ r_{\rm t}$. This maximum
arises from the combination of the fact that smaller periaston
distances result in larger tidal torques and increased mass
loss. With a larger core, the planet is less distorted and at
close separations the mass loss is also suppressed
(Section~\ref{sec:massloss}), so at further separations the tidal
torque is reduced and the peak shifts toward closer separations.

\subsubsection{Mass Loss and Its Asymmetry}
\label{sec:massloss}

\begin{figure*}
  \centering
  \includegraphics[width=\linewidth,clip=true]{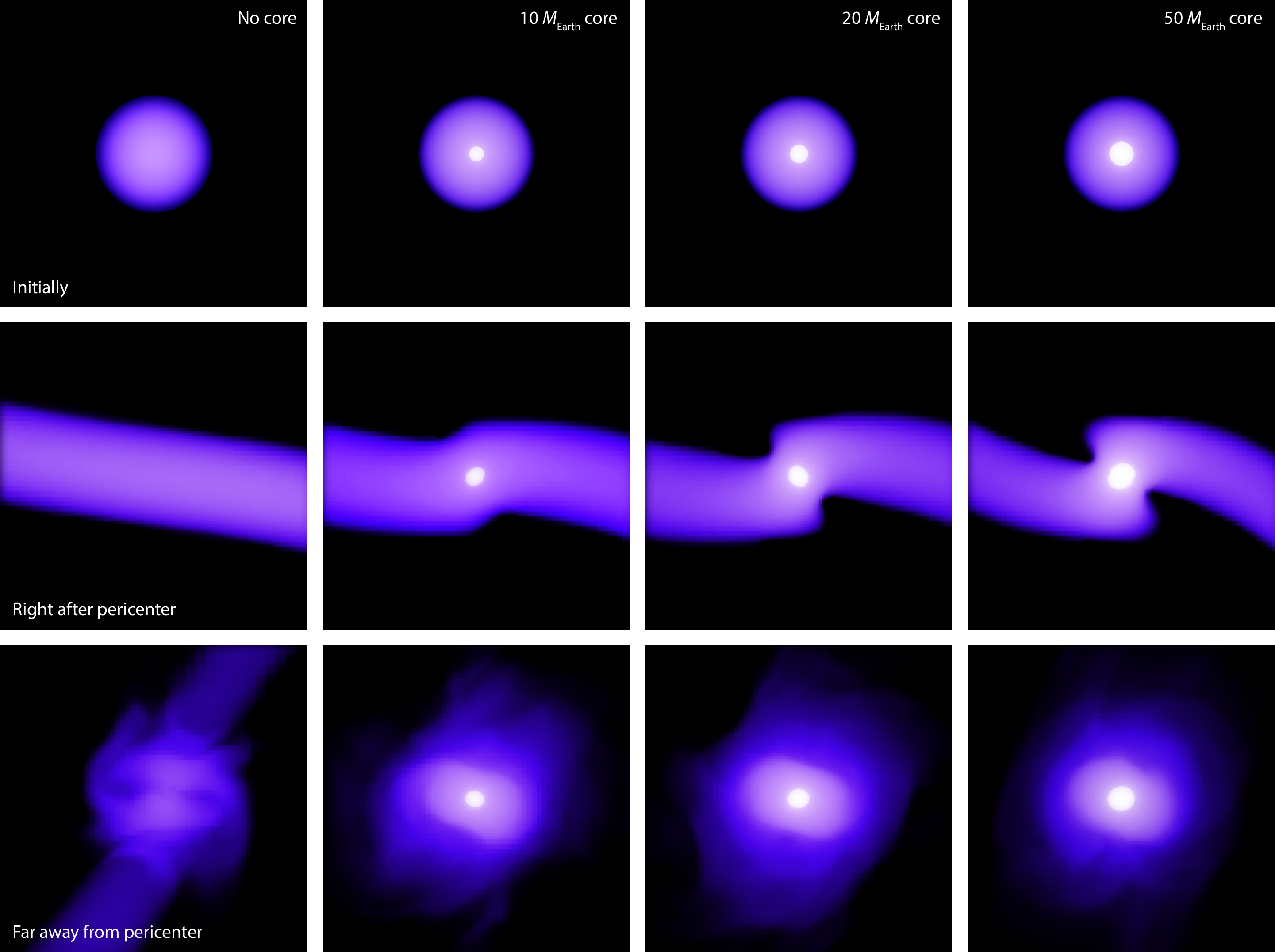}
\caption{Snapshots from disruption simulations of Jupiter-mass
giant planets with different core masses, modeled as a
dual-layered composite polytropes. Four simulations are shown
above (one per column), with the only difference between the
simulations being the mass of the planet's core (as labeled).
The top row shows the initial structure of the planet prior to
being tidally perturbed, the middle row corresponds to the state
of the planet shortly after pericenter, and the last row shows
the planet many dynamical timescales after pericenter. The
planet comes within 1.2 $r_{\rm t}$ of its parent star in each
of the simulations, and is initialized with two separate
components: A core with a stiff gamma-law EOS ($\gamma_{\rm 1} =
3$, shown in white), and an envelope with a softer equation of
state ($\gamma_{\rm 2} = 2$, shown in purple).}
  \label{fig:corecmp}
\end{figure*}

Planets lose mass as a consequence of intense tidal
perturbation, especially during close encounters ($r_{\rm p}<2.0
\ r_{\rm t}$) as illustrated in Figure~\ref{fig:snapshot}. This
mass loss is not symmetric and, as suggested by FRW and GRL, is
responsible for the observed change in $E_{\rm orb}$. The
resultant kick from asymmetric mass loss will be discussed in
detail in Section~\ref{sec:emrelation}. The fraction of mass
unbound from the planet $\Delta m$ in each run is plotted in the
left panel of Figure~\ref{fig:massloss}. The results show that
for encounters with $r_{\rm p} \geq 2.0 \ r_{\rm t}$, tides
raised by the star are too weak to shed any noticeable amount of
mass from the planet, irrespective on its internal structure.
However, in the mass-shedding regime ($r_{\rm p} < 2.0 \ r_{\rm
t}$), the discrepancies between different models are rather
prominent. Although the presence of a core does not alter the
total mass and radius of the planet, planets with cores heavier
than 20 $M_\earth$ lose significantly less mass than those
without cores\footnote[3]{Note that the SPH simulation seems to
underestimate the mass loss in all destructive cases, and the
changes in orbital energy for the deepest encounters as shown in
the left panel of Figure~\ref{fig:EorbJspin}.}. In the case of a
$50 \ M_\earth$ core (corresponding to 15\% of the total mass),
the planet can retain more than half of its mass even for the
deepest encounter we calculated in this work, with a periastron
separation of only $1.15 \ r_{\rm t}$. Note that the stellar
radius imposes a lower limit on the planet's minimum periastron
approach distance; i.e., the sum of star's and planet's radii
$R_*+R_{\rm P}\simeq 1.0995 \ R_\sun \simeq 1.105 \ r_{\rm t}$,
assuming the host star is Sun-like and has a radius of $1 \ R_\sun$.

As a result of the local strength of the tidal field being
proportional to $r^{-3}$, $\Delta m_1$, the fraction of mass
lost through $L_1$, is always greater than $\Delta m_2$, the
fraction of mass lost through $L_2$ ($\Delta m = \Delta m_1 +
\Delta m_2$). We plot the ratio of the mass lost in the two
tidal streams $\Delta m_2 / \Delta m_1$ as a function of the
periastron separation in the right panel of
Figure~\ref{fig:massloss}. We confirm the change of asymmetry of
mass loss as first noted by FRW, in which the inner stream
dominates mass loss at large separations where the total mass
loss is small, while $\Delta m_2 / \Delta m_1$ approaches unity
(but is still less than one) at smaller separations where the
total mass loss is significant. This is also reflected in the
morphological difference between disrupted planets illustrated
in Figure~\ref{fig:snapshot} by the two trajectories $r_{\rm
p}/r_{\rm t}=1.2$ and $r_{\rm p}/r_{\rm t}=1.8$. All the models
with different core masses generally conform to this trend,
albeit models without cores seem to deviate from models with
cores for both small and large values of $r_{\rm p}/r_{\rm t}$,
with the change in energy increasing dramatically for the former
and saturating at a fixed value for the latter. Note that the
mass loss difference may be modified by the magnitude and
orientation of incoming planets' spin if the spin frequency is a
significant fraction of that for break up, with the spin
potentially being accumulated over prior encounters with the
star (Figure \ref{fig:EorbJspin}).

\subsubsection{Core Mass and Survivability}
Planets with cores not only lose less mass, but also maintain
their internal structures more effectively than their coreless
counterparts after the disruption has occurred.
Figure~\ref{fig:corecmp} shows density profiles of various
planet models as they are torn apart near their pericenters
(middle row), and when the remnants are relaxed after many
dynamical timescales (bottom row). Without a core (first
column), the planet is easily shredded, resulting in a long
tidal stream that eventually coalesces into a weakly self-bound
remnant. The envelope of a planet with a core is still
significantly disturbed at pericenter, but the core itself is
only weakly affected (second through fourth columns). This
results in a core-envelope interface that is well-preserved
after the encounter, with planets maintaining a larger fraction
of their original structure for progressively larger core
masses.

\begin{figure*}[tb]
  \centering
  \includegraphics[width=\linewidth,clip=true]{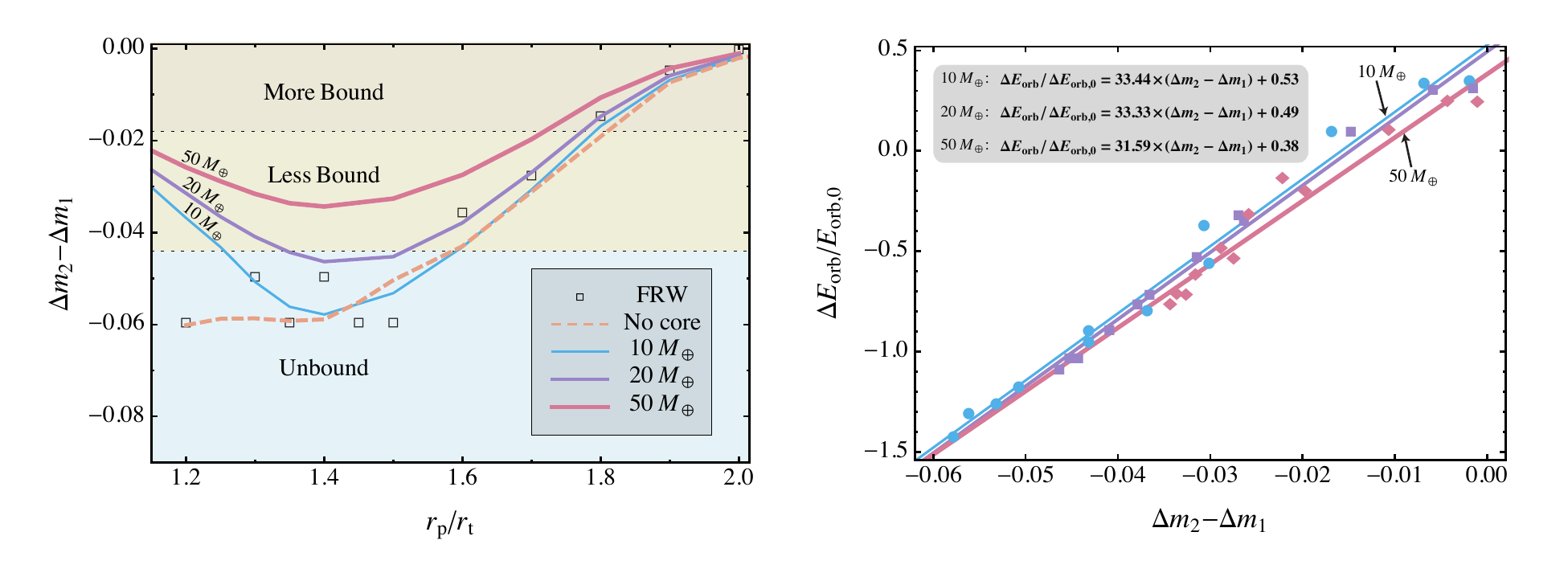}
\caption{The left panel shows the difference between the two
normalized mass loss $\Delta m_2 -\Delta m_1$ as a function of
periastron distance scaled by the tidal radius $r_{\rm p} /
r_{\rm t}$. The color bands have the same meaning as in the left
panel of Figure~\ref{fig:EorbJspin}. The correlation between the
mass loss difference $\Delta m_2 -\Delta m_1$ and changes in
orbital energy scaled to the initial specific orbital energy
$\Delta E_{\rm orb} / E_{\rm orb,0}$ is shown in the right
panel, where the orange triangles, blue points, purple squares
and red diamonds are the simulation data of planets with no
core, $10 \ M_\earth$, $20 \ M_\earth$ and $50 \ M_\earth$ cores
respectively. The blue thin, purple medium, and red thick lines
show the linear least squares fits to the simulations with
cores.}
   \label{fig:emrelation}
\end{figure*}

Cores have long been ignored in hydrodynamic simulations because
they only contribute to a tiny fraction of a planet's total
mass, and have thus been thought to be dynamically
insignificant. However, we show that the core mass is of prime
importance in determining the fate of disrupted planets in the
sense that both the changes in orbit energy and morphology of
the planets are strongly related to their cores. The addition of
cores to models of giant planets may be the key to solving the
overestimated destructiveness of tidal field found in GRL. We
shall discuss the effects of cores in the context of a planet's
adiabatic response to mass loss in Section~\ref{sec:densecore}.

\section{Discussion}\label{sec:dis}
\subsection{The correlation between mass loss and changes in orbital energy}
\label{sec:emrelation}

The left panel of Figure~\ref{fig:emrelation} shows the
normalized mass difference between the two streams $\Delta m_2 -
\Delta m_1$ as a function of the periastron separation for
various planet models. The mass difference $\Delta m_2 - \Delta
m_1$ can be related to the total mass loss $\Delta m$ and the
mass loss ratio $\Delta m_2 / \Delta m_1$ through a simple
relation \begin{equation} \label{equ:massloss} \Delta m_2 -
\Delta m_1=- \Delta m \left(\frac{2}{\Delta m_2 / \Delta m_1+1}
-1 \right). \end{equation} Because $\Delta m$ is a monotonically
decreasing function of $r_{\rm p}/r_{\rm t}$, while the term in
the parentheses on the right hand side of equation
(\ref{equ:massloss}) is a monotonically increasing function of
$r_{\rm p}/r_{\rm t}$, the mass difference maximizes (i.e.
$\Delta m_2 - \Delta m_1$ becomes most negative) for planets
with cores when $r_{\rm p}/r_{\rm t} \simeq 1.4$. One may notice
that the dependence of $\Delta m_2 - \Delta m_1$ on periastron
separation is very similar to that of $E_{\rm orb} / E_{\rm
orb,0}$. Indeed, we find the change in specific orbital energy
scaled to the initial specific orbital energy $\Delta E_{\rm
orb} / E_{\rm orb,0}$ linearly correlates with the normalized
mass difference $\Delta m_2 - \Delta m_1$, where $\Delta E_{\rm
orb}=E_{\rm orb}-E_{\rm orb,0}$ is the change in specific
orbital energy after the tidal disruption (Fitting formulas are
provided in Figure~\ref{fig:emrelation} for reference). Thus,
one may use the mass difference to determine whether a planet's
final orbit is more bound or less bound, as illustrated in
color-shaded regions in the left panel of
Figure~\ref{fig:emrelation}.

What underlies this linear relation is energy conservation.
Material stripped from the planet with negative binding energy
becomes bound to the host star and forms the inner tidal stream,
while that with positive binding energy becomes unbound to the
system and forms the outer tidal stream. Not surprisingly, the
energy deposited into the inner tidal stream is always greater
than that deposited into the outer stream due to the asymmetric
tidal forces, with the degree of asymmetry depending on the
ratio between the star and the planet's masses. As a result, the
change in planet orbital energy reflects the binding energy
difference between the two streams as the total system's energy
must be conserved. For all cases with mass loss, the net energy
exchange between the two tidal streams is negative, resulting in
a positive change in the planet's orbital energy. Thus, the
planet becomes less bound (or unbound) to the star assuming
any other form of energy exchange can be neglected. This
assumption holds for all deep encounters with large mass loss.

However, two additional sinks of energy exist: The energy stored
within the planet's normal modes of oscillation $E_{\rm osc}$,
and the energy associated with the planet's final spin $E_{\rm
spin}$. The sum of these two terms cannot exceed the planet's
own self-binding energy, $E_{\rm bind} \simeq G M_{\rm
P}^{2}/R_{\rm P}$, and this reflects the maximum negative change
in orbital energy that can be achieved in a single passage. As
we show in the right panel of Figure~\ref{fig:EorbJspin},
planets gain most spin angular momentum at separations where the
mass loss is relative small. In other words, the rotational and
oscillatory kinetic energies saturate for encounters in which
little mass is lost, and thus cannot aid in retaining the planet
for deeper encounters.

Because the asymmetric mass loss could kick the planet into a
less bound orbit, one may conclude that the continuous positive
change in the planet's orbital energy could lead to a planetary
ejection after several close-in passages. However, this
statement overlooks several crucial facts. First, the planet
loses a significant fraction of its mass during the first
passage, leading to an increase in the mass ratio of subsequent
encounters. Second, while the planet's envelope becomes
progressively less dense after each encounter, the core remains
intact. As a result, the thrust provided by the loss of the
envelope becomes less effective as its mass decreases (the
effects of the core on the survivability of the planet will be
discussed in Section~\ref{sec:adiabatic}). As the envelope is
depleted, the effective tidal radius $r_{\rm t}$ increases.
Consequently, in the following passages the planet may have
$r_{\rm p}/r_{\rm t}$ values close to or even less than unity,
and the two tidal streams produced by subsequent encounters will
be more equal in mass, as suggested by our simulations (see
Figure~\ref{fig:emrelation}). As the ratio of core mass to total
mass is enhanced, this results in a larger specific self-binding
energy for material close to the core-envelope interface, and if
this material is perturbed, it can absorb a larger fraction of
the planet's orbital energy. As a result, the specific orbital
energy may become more negative than in the case where no core
is present. We also did not consider the possibility that
planets are scattered from distances inside their parent stars'
respective snow lines. In those cases, the energies required to
unbind the planets are much larger than the energy exchanged in
the encounter, and planets are more likely to be bound to the
star, even after multiple passages.

\subsection{Adiabatic Responses of Mass-losing Composite Polytropes}
\label{sec:adiabatic}

When a planet loses mass on a timescale faster than the
Kelvin-Helmholtz time but slower than the dynamical time, the
structure of the planet will evolve adiabatically so that the
entropy as a function of interior mass is approximately
conserved \citep{Dai:2011fk}. \citet{Hjellming:1987ys} used
composite polytropic stellar models ($P\propto
\rho^\gamma\propto \rho^{1+{1\over n}}$) to explore the
stability of this adiabatic process. We have modified their
formalism slightly by taking into account a distinct chemical
composition between the core and envelope, $\mu_1 = 4 \ \mu_2$.
We also consider a different combination of polytropic indices,
$n_1=0.01$ and $n_2=1$. Note here we choose the more realistic
value of $n_{\rm 1} = 0.01$ instead of $n_{\rm 1} = 0.5$ used in
our hydrodynamic simulations, because we want to investigate the
extreme case in which the entire envelope is shed. A stiff EOS
is required in order to capture the core's incompressible
response to pressure deformations. The reader is refer to
Appendix~\ref{app:adiabatic} for a detailed description of our
adiabatic response to mass loss model for composite polytropes.

\begin{figure}[tb]
  \centering
   \includegraphics[width=\linewidth,clip=true]{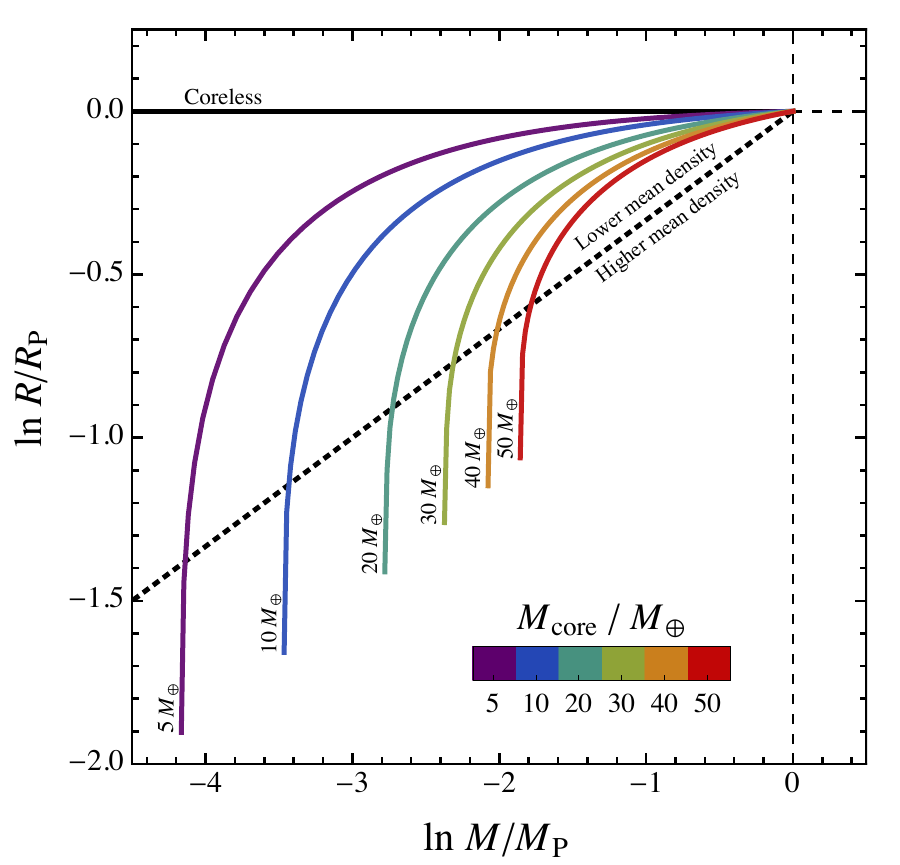}
\caption{Adiabatic response curves for composite polytropes for
varying degrees of mass loss. All curves start from the origin
(zero mass loss). A negative value indicates either a decrease
in mass (x-axis) or a shrinkage of the planet's radius (y-axis).
The top horizontal thick black line shows the evolution for a
single-layered $n=1$ polytrope model. From left to right the
solid lines in colors are the response curves of composite
polytropes with 5, 10, 20, 30, 40, and $50 \ M_\earth$ cores,
respectively. For all the composite polytrope models, we assume
$n_{\rm 1} = 0.01$ and $n_{\rm 2} = 1$, and set the two layers
to have different molecular weights, with $\mu_{\rm 1} = 4 \
\mu_{\rm 2}$. The dotted line (whose slope is 1/3) illustrates
the response required to have no noticeable change in the
planet's mean density. Below this line, composite polytropes
become more dense in response to further mass loss.}
  \label{fig:adiabatic}
\end{figure}

In contrast to the single-layered $n=1$ polytrope, which has a
constant radius\footnote[5]{As noted by
\citet{Hjellming:1987ys}, if the relation $\gamma=1+1/n$ is
valid, the radius change of a single-layered polytrope comforms
to a simple function: $R=\omega_0^{(1-n)/2(n+1)} R_0$, where
$\omega_0$ is a quantity describing the mass loss. In the case
of $n=1$, the dependence of $\omega_0$ vanishes.}, composite
polytropes with a small $n_{\rm 1}$ always contract as they lose
mass. As illustrated in Figure~\ref{fig:adiabatic}, the planet's
contraction rate is observed to increase as the core mass
fraction increases. The way that a planet evolves and ends up in
such a tidally mass-losing environment depends upon how its mean
density (or size) changes as it losses mass. The dotted line in
Figure~\ref{fig:adiabatic} denotes the boundary below which the
adiabatic response will lead to an increase in the mean density
of the stripped planet and, as a result, the stripped object
will be less vulnerable to tidal disruption. However, during the
tidal circularization process the planet's periastron distance
$r_{\rm p}$ will increase due to conservation of orbital angular
momentum. Once $r_{\rm p}$ becomes sufficiently large (say about
$2.5\ r_{\rm t}$) such that stellar tides can no longer strip
significant mass from the planet, a lower mean density could
still arise as a result of tidal heating.

Figure~\ref{fig:adiabatic} can help explain the dependence of
our results on the assumed polytropic index of the core. The
computed coreless model is equivalent to the composite polytrope
with an $n_1=1$ core (ignoring of course that there is no
density jump at the core-envelope interface in this model).
Thus, one can qualitatively infer that the adiabatic response of
a core's model with $0.01<n_1<1$ would lie between the two model
extremes represented in Figure~\ref{fig:adiabatic}. By doing so
we see that only a very small discrepancy in the adiabatic
responses of the extreme models depicted in
Figure~\ref{fig:adiabatic} can be observed until as much as half
of the mass of the planet is lost ($M/M_{\rm P} \sim 0.5$). This
gives credence to the idea that our results are rather
insensitive to $n_1$, because the overall adiabatic response is
mainly determined by the remaining envelope, which is at least
an order of magnitude more massive than a 10 $M_\Earth$ core.
This is consistent with our additional hydrodynamic simulation,
where no dramatic difference between models with $n_1=0.5$ and
$n_1=0.01$ was found after a single close encounter.

As the mass loss increases, the mass of the core becomes
progressively more important in determining the adiabatic
response of the planet. When the mass of the remaining envelope
becomes comparable to the mass of the core, the overall response
of the composite polytrope to mass loss is primarily dictated by
the core. At this stage, models with different $n_1$ behave very
differently. For these extremes cases, a small $n_1=0.01$ is
more suitable to model the incompressible behavior of a rocky
core.

\subsection{The Role of Dense Cores}
\label{sec:densecore}
Our hydrodynamical simulations have demonstrated that with
larger cores, planets retain a greater fraction of their
original envelope (see Figure~\ref{fig:corecmp} for comparison),
which also means that the difference between the mass lost from
the near- and far-sides of the planet is reduced, and therefore
the planet has a greater chance to be bound to the host star.
This result is somewhat surprising as the cores contain only a
small fraction of the planet's total mass, and have been
regarded as being dynamically unimportant (e.g. FRW and GRL).
Previous studies which have attempted to determine the final
fate of disrupted hot Jupiters usually ignore the complexity of
the planet's interior structure.

Recently, \citet{Remus:2012uq} investigated the dissipative
equilibrium tide in gas giant planets by taking into account the
existence of viscoelastic cores. While approaches similar to
theirs can predict the amount of energy deposited into a
planet's interior by an external tidal perturber, such
formalisms fail for disruptive encounters in which non-linear
dynamical effects dominate and a significant fraction of mass is
removed from the planet. To approximate dynamical mass loss, we
calculate the adiabatic response of composite polytropes.
Cooling is ignored in our analysis as its time scale is
significantly longer than the dynamical time scale
\citep{Bodenheimer:2001qf}, but we note that it can be important
in determining the planet's final structure once mass loss
ceases \citep{Fortney:2007fk}.

The $n=1$ single-layered polytrope, which corresponds to the
coreless gas giant planets, does not change its radius when
losing mass adiabatically, resulting in a decrease of the
average density. By contrast, the extremely incompressible cores
of composite polytropes are weakly affected by the perturbation,
imposing an almost constant inner boundary condition for the
envelope, and resulting in an increase in density when the
core's gravity dominates (see Figure~\ref{fig:adiabatic}). This
phenomenon helps to explain the different amounts of mass lost
in the two cases. Each time the single-layered polytrope loses
some mass, the specific gravitational self-binding energy
decreases, leading to a more tidally-vulnerable structure. As a
result, GRL found that coreless planets are always destroyed
after several passages even if the initial periastron is fairly
distant (the lower limit is $2.7 \ r_{\rm t}$). The composite
polytropes, on the other hand, maintain a constant gravitational
potential well in their centers, which continuously resists the
stellar tidal force. Being invulnerable to tidal disruption
themselves, the cores survive, retaining some fraction of the
original envelope \citep{MacLeod:2012fk}. This results in a core
mass fraction that is significantly larger than that of the
original planet.
 
We should emphasize that although we focus on Jupiter-mass
planets in this work, the scenario presented here can apply to
giant planets of different masses, as long as they are
characterized by a similar dual-layered structure. There are
several known gas giant planets with very large average
densities and enhanced metallicities \citep{Bakos:2011uq}.
CoRoT-20 b, for instance, with a core mass fraction between 50\%
and 77\% of its total mass, orbits a G-type star on an eccentric
orbit ($e=0.562$) \citep{Deleuil:2012fk}. It is not clear yet
how these metal-rich giant planets are formed. Tidal disruption
might be an explanation. In this scenario, these planets were
more massive prior to disruption, with a more typical fraction
of heavy elements concentrated in their cores. After the loss of
the envelope, the planet's average properties, including
metallicity, become more representative of the core's initial
properties. Measurements of the metallicity enhancement,
combined with measurements of the planet's orbital properties,
may enable one to infer the planet's original mass.

\subsection{Demographics of the Surviving Tidally Disrupted Giant Planet Population}
\label{sec:population}

\begin{figure*}[t]
  \centering
  \includegraphics[width=0.85\linewidth,clip=true]{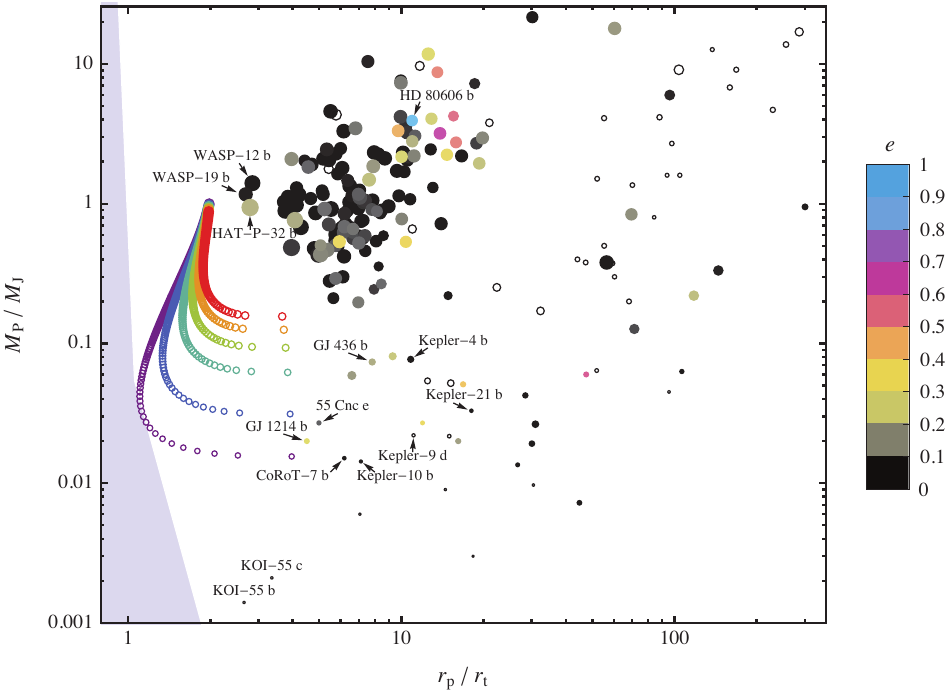}
\caption{Planet mass $M_{\rm P}$ versus pericenter distance
$r_{\rm p}$ scaled by the tidal radius $r_{\rm t}$. The sample
is composed of planets with known mass and radius and a known
stellar mass, from which the tidal radius $r_{\rm t}$ for each
planet is then computed. The filled points are color-coded by
the planet's eccentricity. For the planets with unknown
eccentricity, $e = 0$ is assumed. The open points illustrate the
adiabatic evolution of a Jupiter-mass planet experiencing mass
loss with different core masses, with the color coding being the
same as in Figure~\ref{fig:adiabatic}. The blue shaded region in
Figure~\ref{fig:exoplanet} is plotted to illustrate how
$R_{\odot}$ compares with $r_{\rm t}$ as the planet's mass is
varied, with the relationship being linearly interpolated
between the Sun-Earth, Sun-Neptune and Sun-Jupiter cases. All
exoplanet data was taken from the Extrasolar Planets
Encyclopaedia (\href{http://exoplanet.eu}{http://exoplanet.eu/})
on June 14, 2012, the sample contains 217 planets.}
  \label{fig:exoplanet}
\end{figure*}

\begin{deluxetable*}{lccccccccc}
    \tablecolumns{10}
    \tablewidth{0pc}
    \tablecaption{A sample of extremely close-in super-Earths and Neptune-size planets\tablenotemark{a}}
    \tablehead{\colhead{Name} & \colhead{$M_{\rm P}$} & \colhead{$R_{\rm P}$}  & \colhead{$a$} & \colhead{Period} & \colhead{$e$} &  \colhead{$M_*$} & \colhead{$\rho_{\small \rm P}$} & \colhead{$r_{\rm p}/r_{\rm t}$\tablenotemark{b}} & \colhead{$r_{\rm p,0}/r_{\rm t,0}$}\\
   & \colhead{($M_\earth$)}  & \colhead{($R_\earth$)}  & \colhead{(AU)} & \colhead{(days)} & & \colhead{($M_\odot$)} & \colhead{($\rho_{\small \rm J}$)} &}
    \startdata
     55 Cnc e & 8.58 & 2.11 & 0.0156 & 0.74 & 0.06 & 0.91 & 3.70 & 4.99 & 1.71 \\
     Kepler-10 b & 4.55 & 1.38 & 0.0168 & 0.84 & 0 & 0.90 & 6.98 & 7.11 & 1.86 \\
     CoRoT-7 b & 4.80 & 1.63 & 0.0172 &  0.85 & 0 & 0.93 & 4.47 & 6.18 & 1.88 \\
     GJ 1214 b & 6.36 & 2.66 & 0.014 & 1.58 & 0.27 & 0.15 & 1.36 & 4.51 & 2.58 \\
     Kepler-9 d & 7.00 & 1.60 & 0.0273 & 1.59 & 0 & 1 & 6.93 & 11.08 & 2.91 \\
     GJ 436 b & 23.42 & 3.96 & 0.029 & 2.64 & 0.15 & 0.45 & 1.52 & 7.82 & 3.92 \\
     Kepler-21 b & 10.49 & 1.58 & 0.0425 & 2.79 & 0 & 1.34 & 10.49 & 18.05 & 4.11 \\
     Kepler-4 b & 24.47 & 3.88 & 0.0456 & 3.21 & 0 & 1.22 & 1.69 & 10.82 & 4.54
    \enddata
    \tablenotetext{a}{All data was taken from the Extrasolar Planets Encyclop\ae dia (\href{http://exoplanet.eu}{http://exoplanet.eu/}) on June 14, 2012.}
    \tablenotetext{b}{The current $r_{\rm p}/r_{\rm t}$ values are also plotted in Figure~\ref{fig:exoplanet}.}
    \label{tab:candidates}
\end{deluxetable*}

\subsubsection{A Census of Exoplanets with Known Mass and Radius}
\label{subsec:exoplanets}
Despite the rapid pace of exoplanet discovery and influx of 
dynamical data, information on the structure of exoplanets
is still limited. Both mass and radius have been determined
for several hot Jupiters and super Neptunes.  Although there is
essentially no direct information on their internal structure,
the average density of these planets is likely to be correlated 
with the presence of cores \citep{Miller:2011kx}.  

In order to search for some clues on the role of tidal
disruption during their orbital circularization process, we show
a sample of exoplanets with known planetary radii
$R_{\rm P}$ and masses $M_{\rm P}$ and known stellar masses
$M_*$ in Figure~\ref{fig:exoplanet}, where the distribution of
the planet's mass as a function of its pericenter distance
scaled by its tidal radius
\begin{equation}
  \frac{r_{\rm p}}{r_{\rm t}} = \frac{a(1-e)}{(M_*/M_{\rm P})^{1/3}R_{\rm P}}
\end{equation}
is plotted. The color-coding of the filled dots denotes
eccentricity $e$, and the open black circles represent planets
with unknown eccentricity, where we have assumed $e=0$ for our
subsequent calculations. The size of the symbols is
representative of the planet's physical size (not drawn to
scale). The most eccentric planet in our sample is HD 80606b
(the current record-holder HD 20782 b with $e=0.97$ is excluded
because the size of this planet has not been measured). Note
that there is a non-trivial fraction of planets with
substantially large eccentricities. Moreover, both the radial
velocity and transit surveys are biased against the detection of
highly-eccentric planets \citep{Socrates:2012fk}, as such, the
fraction of these planets is likely under-represented. It is
also important to notice that the most eccentric planets are
found at larger separations, though this may be enhanced by
detection bias. This is consistent with the scenario that
dynamical processes such as planet-planet scattering
\citep{Rasio:1996kx, Chatterjee:2008gd, Ford:2008vn} or the
Kozai mechanism \citep{Kozai:1962fk, Takeda:2005kx, Naoz:2011vn,
Nagasawa:2011dp} lead to the excitement of a planet's
eccentricity, while tidal dissipation can damp the planet's
eccentricity in the vicinity of the star.

\citet{Zhou:2007fk} and \citet{Juric:2008uq} studied the eccentricity
distribution of dynamically relaxed exoplanetary systems, which can be
described by a Rayleigh distribution
\begin{equation} 
dN =\frac{e}{\sigma_e^2} {\rm exp} \left( \frac{-e^2}{2\sigma_e^2}\right)de, 
\end{equation} 
where $\sigma_e = 0.3$. This distribution has a small, but
non-negligible fraction of planets with eccentricities close to
unity, with the number of planets having $e = 0.997$ being only
a factor of about 2.5 smaller than $e=0.9$. Given the observed
number of eccentric planets with $0.9 < e < 1$, we thus expect
some super-eccentric planets with $e > 0.997$. For $e=0.997$,
the periastron distance for a planet scattered from the snow
line given a Sun-like parent star would be $r_{\rm p}=a (1-e) =
0.0075 \ {\rm AU} = 1.5 \ R_\odot$. For these small separations, the
budgeting of the planet's orbital energy during a tidal
encounter must account for mass loss, as the final orbital
energy is strongly correlated with the properties of the ejected
mass (Figure~\ref{fig:emrelation}).
 
To determine the evolution of giant planets in the $(M_{\rm P}$,
$r_{\rm p}/r_{\rm t})$ plane, we calculate the tidal radius of
composite polytropes using the structure predicted by the
response to adiabatic mass loss. An initial periastron
separation $r_{\rm p}=2.0 \ r_{\rm t}$ is adopted. We assume that
orbital angular momentum is conserved, and that the planet ends
up in a circular orbit ($a \simeq 2 \ r_{\rm p}$). The tracks in
Figure~\ref{fig:exoplanet} show a Jupiter-mass planet evolving
into a super-Earth or a Neptune analogue (depending on the
initial core mass) as its envelope is continuously removed. The
tracks show that $r_{\rm p}/r_{\rm t}$ first decreases to a
minimum value as the average density of the planet decreases,
but this trend reverses as the importance of the gravitational
influence of the core on the remaining envelope increases
(Figure~\ref{fig:adiabatic}). Note that the adiabatic response
model assumes that mass is slowly removed, and that no
additional energy is injected into the envelope.

Based on the models of adiabatic mass loss, it seems plausible
that giant planets with cores can be transformed into either
super-Earths or Neptune-like planets during their orbital
circularization process. As the adiabatic model only makes
predictions about the structure of the planet, the model alone
is incapable of determining whether the planet would be
circularized or ejected, which depends on how the mass is
removed from the planet.

Depending on the initial mass of the planet and its core, the
final mass and radius of the planetary remnant can vary
drastically (as illustrated in Figure~\ref{fig:exoplanet}).
Planets that lie near the evolutionary end states for a
disrupted Jupiter-like planet are tabulated in
Table~\ref{tab:candidates}. To determine if a planet may have a
tidal disruption origin we compute its initial $r_{\rm
p,0}/r_{\rm t,0}$ value assuming the specific orbital angular
momentum is conserved during the circularization process, so
that
\begin{equation}
 r_{\rm p,0}\simeq a_{\rm f}/2=r_{\rm p}(1+e)=a(1-e^2),
\end{equation}
where $a_{\rm f}$ is the semimajor axis after  the orbit has been
circularized. If the gas giant progenitor has Jupiter's mean
density, the initial tidal radius is then given by
\begin{equation}
r_{\rm t,0}=r_{\rm t}\rho_{\small \rm P}^{1/3},
\end{equation}
where $\rho_{\small \rm P}$ is the presently observed density of
the planet. Among these planets, 55~Cnc~e, Kepler-10~b and
CoRoT-7~b have $r_{\rm p,0}/r_{\rm t,0} \lesssim 2$, which
guarantees that if the planet was initially similar to Jupiter
it would have lost mass on its first encounter with the parent
star (FRW; GRL). For planets with $2 \lesssim r_{\rm p,0}/r_{\rm
t,0} \lesssim 2.7$ (e.g. GJ~1214~b), prolonged tidal effects
over many orbits may still lead to significant mass loss (GRL).
Planets with $r_{\rm p,0}/r_{\rm t,0} \gtrsim 2.7$ are unlikely
to have formed as a result of the tidal disruption of a
Jupiter-like planet, unless they were significantly less dense
than Jupiter, which is possible at the time of scattering as may
not have had sufficient time to cool \citep{Fortney:2007fk}.

Of the transiting super-Earths with known masses, only a few
presently lie within a few tidal radii of their host stars. This
may indicate that the conditions necessary to generate such
planets via tidal disruption are uncommonly realized in nature.
However, the sample is highly biased against low-mass planets,
simply because we need transit surveys to determine the planet's
size. In addition, the eccentricities for many low-mass planets
are poorly constrained, and as a result, their periastron
separations may have been overestimated. This highlights the
importance of conducting a survey that is capable of detecting
close-in, low-mass planets, such as the {\it Kepler} mission.

\begin{figure}[tb]
  \centering
  \includegraphics[width=\linewidth,clip=true]{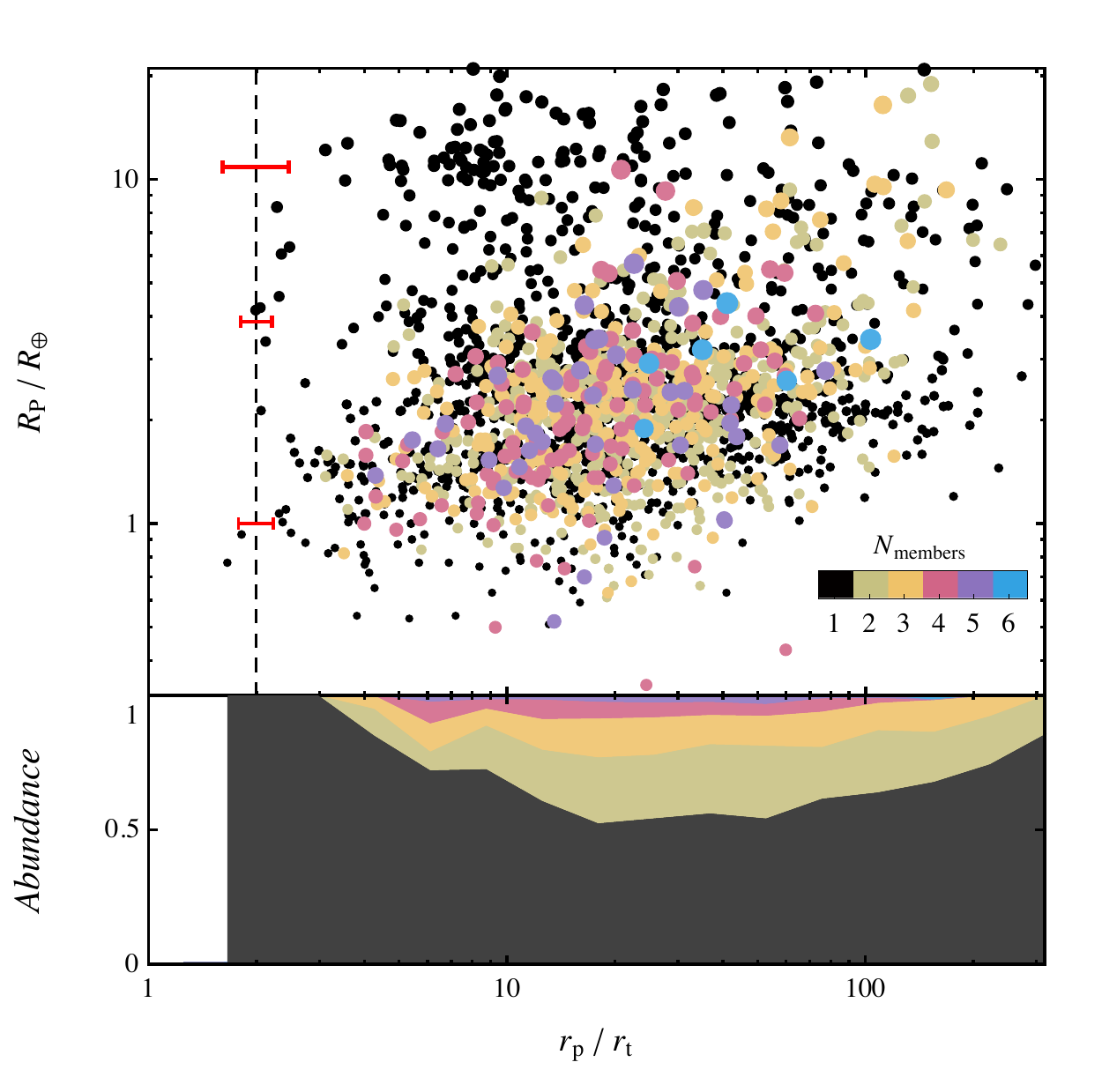}
\caption{The upper panel shows the radius in Earth radii of {\it
Kepler} planet candidates versus pericenter distance in units of
the tidal radius. The color of each point denotes the
multiplicity of the system in which it resides
\citep{Batalha:2012uq}, while the dashed vertical line indicates
$2 r_{\rm t}$. To determine the tidal radius without explicit
measurements of the mass, the density of each candidate planet
is assumed to be equal to that of a similarly-sized planet in
the solar system. From top to bottom, the red bars represent the
range of $r_{\rm p}/r_{\rm t}$ values at typical sizes of
Jupiter, Neptune and Earth, respectively (see the text for more
details). In the lower panel, a stacked area plot shows the
fraction of single-candidate and multi-candidate systems, with
the color-coding being the same as the upper panel. All data was
take from NASA Exoplanet Archive
(\href{http://exoplanetarchive.ipac.caltech.edu/}{http://exoplanetarchive.ipac.caltech.edu/}).}
  \label{fig:kepler}
\end{figure}

\subsubsection{A Census of Kepler Candidates}
\label{subsec:kepler}
In the upper panel of Figure~\ref{fig:kepler} we plot the
distribution of {\it Kepler} candidate radii as a function of
their periastron separations (taken to be equal to the semimajor
axes) divided by their tidal radii, which we obtain making use
of equation (\ref{equ:tidalradius}). Unfortunately, for most of
the candidates discovered by transit we don't known the
planetary mass because usually these stars too faint to do
radial velocity measurements except for those in multiple
systems where transit time variation (TTV) measurement is
possible (one of the successful measurement is the Kepler-11
system done by \citet{Lissauer:2011el}). To estimate the mass of
each candidate, we use the density of planets in our solar
system: For candidates whose sizes are equal or larger than
Jupiter, we take the geometric mean of Jupiter's and Saturn's
densities ($\rho_{\small \rm P}=0.96$ g cm$^{-3}$), for Neptune
size candidates we use Neptune's and Uranus' densities
($\rho_{\small \rm P}=1.51$ g cm$^{-3}$), and for Earth and
sub-Earth size candidates we use Earth's and Mars' densities
($\rho_{\small \rm P}=4.66$ g cm$^{-3}$). We linearly
interpolate between the three cases for planets of intermediate
radii. The red bars in Figure~\ref{fig:exoplanet} show the range
of $r_{\rm p}/r_{\rm t}$ values calculated with two limiting
densities at each typical size. While we have assumed that the
semimajor axis $r_{\rm p} = a$, \citet{Kane:2012ys} note that
some of the candidates may still have large eccentricities.
However, as illustrated in Figure~\ref{fig:exoplanet}, most
planets within 0.04 AU are expected to have $e \simeq 0$.

\begin{deluxetable}{lcccc}
    \tablecolumns{5}
    \tablecaption{A sample of close-in Kepler candidates}
    \tablehead{\colhead {KOI Name} & \colhead{$R_{\rm P}$} &\colhead{$M_{\rm P}$\tablenotemark{a}} &\colhead{$M_*$} &  \colhead{$r_{\rm p}/r_{\rm t}$}   \\ 
    & \colhead{($R_\earth$)} & \colhead{($M_{\oplus}$)}  &  \colhead{($M_{\sun}$)} & } 
    \startdata
     799.01 & 6.07 & 52.10 & 0.83 & 2.35 \\
     861.01 & 1.77 & 3.01 & 0.79 & 2.49 \\
     928.01 & 2.56 & 6.64 & 0.91 & 2.74 \\
    1150.01 & 1.10 & 1.05 & 1.02 & 2.42 \\
	1164.01 & 0.77 & 0.39 & 0.55 & 1.66 \\
	1187.01 & 3.38 & 11.95 & 0.80 & 2.12 \\
	1285.01 & 6.36 & 58.76 & 0.85 & 2.48 \\
	1419.01 & 8.31 & 115.89 & 0.95 & 2.28 \\
	1442.01 & 1.36 & 1.69 & 1.07 & 2.78 \\
	1459.01 & 4.17 & 19.42 & 0.58 & 1.99 \\
	1502.01 & 2.13 & 4.48 & 0.80 & 2.06 \\
	1510.01 & 1.50 & 2.09 & 0.77 & 2.64 \\
	1688.01 & 0.93 & 0.68 & 0.73 & 1.82 \\
	1812.01 & 4.57 & 24.78 & 0.93 & 2.31 \\
	2233.01 & 1.61 & 2.44 & 0.88 & 2.55 \\
	2266.01 & 1.63 & 2.51 & 0.71 & 2.99 \\
	2306.01 & 0.94 & 0.70 & 0.54 & 2.50 \\
	2347.01 & 1.01 & 0.86 & 0.55 & 2.37 \\
	2404.01 & 1.50 & 2.09 & 0.91 & 2.91 \\
	2492.01 & 0.88 & 0.58 & 1.00 & 2.74 \\
	2542.01 & 1.07 & 0.99 & 0.51 & 2.31 \\
	2573.01 & 4.24 & 20.31 & 0.64 & 2.05
    \enddata
    \tablenotetext{a}{Planet density is estimated based on the densities of planets of our own solar system (see the text for details).}
    \label{tab:kepler}
\end{deluxetable}

Some candidates with sizes smaller than Jupiter have present-day
orbits with $a \sim 2 \ r_{\rm t}$ (indicated by the vertical
dashed line in the upper panel of Figure~\ref{fig:kepler}), and
thus are potentially the surviving remnants of tidally
circularized giant planets with cores. Table~\ref{tab:kepler}
lists all the {\it Kepler} candidates with sub-Jupiter sizes and
$r_{\rm p}/r_{\rm t}<3$ for reference.

Among {\it Kepler} cataloged stars, 20\% of them have multiple
planet candidates \citep{Batalha:2012uq}. The points in the
upper panel of Figure~\ref{fig:kepler} are colored to display
the multiplicity of the system. We do not attempt to
statistically study the differences between singles and multiple
systems in detail, here we simply count numbers of planet
candidates for single and multiple systems and compare their
relative abundances. The result is shown in the lower panel of
Figure~\ref{fig:kepler}, where again the color denotes
multiplicity. Intriguingly, the candidates found in multiple
systems tend to lie further from their parent stars, and none of
the candidates listed in Table~\ref{tab:kepler} are observed to
belong to a multiple candidate system (although they might have
distant, unobserved siblings). Planets in compact multiple
systems are thought to be formed through orbital migration, such
as the Kepler 11 system, which hosts six planets. The fact that
the distribution of the semi-major axis within a few tidal radii
differs for single-candidate versus multiple-candidate systems
suggests that close-in candidates may have instead formed via
dynamical interaction. However, many of the single-candidate
systems may be false-positives, whereas the false-positive
fraction is much reduced for the multiple-candidate systems
\citep{Lissauer:2012uq}. But even under the pessimistic
assumption that half of the observed close-in candidates are
false-positives in Table~\ref{tab:kepler}, there are still
around a dozen candidates in the currently available sample that
are close enough to their parent stars to have had a strong
tidal encounter in which much of their original mass was lost.
The conformation of these candidates as true planets opens the
possibility that they have undergone a radical transformation.

In principle, difference in formation histories may be used to
distinguish between the residual cores of gas giant planets and
the failed cores which failed to accrete significant amounts of
gas. For example, under high pressure, metals may react with
hydrogen to produce metallic hydrides such as iron hydride
\citep[][Q. Williams 2012, private
communication]{Badding:1991ve}, which would reduce the core's
mean density from that of a pure metal composition. These
reactions are not expected to happen in failed cores and might
be the only discernible signature as the difference in planetary
densities between these two scenarios may be too small to be
observable.

\section{Summary}\label{sec:summary}
\citet{Nayakshin:2011k} studied the scenario that the tidal
disruption of giant planets occurs during the migration phase of
planetary formation. Under this scenario, the planets migrate
inwards faster than their cooling timescales, and are disrupted
before they can contract and become more resistant to tides.
However, if these planets underwent disk migration, all close-in
planets should have eccentricities near zero. But as there seems
to be an eccentricity gradient, with non-zero eccentricities
being sustained for planets just exterior to their present-day
tidal radii, dynamical interactions followed by dissipative
processes that depend on the distance to the host star offer an
attractive explanation for producing the observed hot
Super-Earths, Neptunes, and Jupiters. While it is unknown if
other effects can produce the observed eccentricity
distribution, scattering events that place planets onto
disruptive orbits are likely to occur in some fraction of
planetary systems.

In this paper, we presented three-dimensional hydrodynamical
simulations of the disruption of giant planets. In contrast to
previous work, we model the planets by including the dense cores
that may exist in the interiors of many (if not most) giant
planets. We show that cores as small as $10M_\earth$ can
increase both the fraction of planets that survive, and the
fraction that remain bound to the host star after a tidal
disruption, and that larger cores make such outcomes more
probable. This is contrary to what has been predicted by
previous simulations in which the giant planets were assumed to
be without cores, where the planets were found to receive large
kicks that would eject them from their host stars, and/or be
destroyed in the process (FRW and GRL). We show that the change
in orbital energy is linearly related to the difference in mass
between the two tidal streams, suggesting that simple energy
conservation arguments are sufficient to explain the observed
post-disruption kicks. We compared our results to the adiabatic
response of composite polytropes to mass loss, and find that
while coreless planets always expand in response to mass loss,
planets with cores contract, allowing them to retain a fraction
of their initial envelopes.

Based on these results, we propose that some gas giant planets
with dense cores could be effectively transformed to a
super-Earth or Neptune-size object after multiple close
encounters. Some of these transformed planets may already exist
within the currently known sample of exoplanets, and are
expected to be small, dense objects that lie close to their
parent stars. The paucity of very close-in exoplanet candidates
in multiple systems found by {\it Kepler} might suggest that the
ordered, gentle migration that typifies most of these systems
may not be universal, and that some systems may evolve via
intense periods of dynamical evolution. One possible signature
of such a dynamical intense event may be an enhancement of the
stellar metallicity as a result of chemical pollution
\citep{Li:2008vn}, or a misalignment between the planet's orbit
and the parent star, as is measured via the Rossiter-MacLaughlin
effect. If it can be determined that some of the observed sample
of close-in Neptunes and super-Earths are relics of this
dynamical history, we may be better equipped to understand the
nature of late-phases of planetary formation.

\acknowledgments
We would like to thank Daniel Fabrycky, Jonathan Fortney,
Tristan Guillot, Morgan MacLeod, Neil Miller, Quentin Williams
and Vivien Parmentier for valuable discussions and perceptive
comments. We would also like to thank the anonymous referee for
thoughtful suggestions which resulted in a greatly improved
paper. The software used in the hydrodynamic simulations was in
part developed by the DOE-supported ASCI/Alliance Center for
Astrophysical Thermonuclear Flashes at the University of
Chicago. Computations were performed on the Laohu computer
cluster at NAOC and the Laozi and Pleiades clusters at UCSC. We
acknowledge support from the David and Lucile Packard
Foundation, NSF grants: AST-0847563 and AST-0908807, NASA
grants: NNX08AL41G and NNX08AM84G, and the NESSF graduate
fellowship (J.F.G.). S.-F.L. acknowledges the support of the
NSFC grant 11073002.

\appendix
\section{Adiabatic Response of Composite Polytropes with Distinct Chemical Compositions}
\label{app:adiabatic}
Adiabatic responses of composite polytropes to mass loss in the
context of binary systems have been investigated by
\citet{Hjellming:1987ys}. They introduced a parameter $w$ to
represent variations in the central pressure, and solved the
Lane-Emden equation in a set of Lagrangian coordinates. We
incorporate their formalism and use separate molecular weights
($\mu_1$ and $\mu_2$) for the core and envelope to represent
their distinct chemical compositions. Following their notations,
the continuity equations (equation (22) and (23) in their work)
become
\begin{equation}
  \label{lagrangian}
  \frac{x_{1}}{\theta_{1}} \frac{d\theta_{1}}{dx_{1}} = \frac{x_{2}}{\theta_{2}} \frac{d\theta_{2}}{dx_{2}},
\end{equation}

\begin{equation}
  x_{1} \theta_{1}^{\frac{n_{1}-3}{6(n_{1}+1)}} \mu_{1}^{-2/3} = x_{2} \theta_{2}^{\frac{n_{2}-3}{6(n_{2}+1)}} \mu_{2}^{-2/3}.
\end{equation}

We have following conditions at the core-envelope interface:
\begin{equation}
  x_{2} = x_{1},
\end{equation}

\begin{equation}
  \theta_{2} = \theta_{1}^\lambda \left(\frac{\mu_{1}}{\mu_{2}}\right)^{-\frac{2}{3}},
\end{equation}

\begin{equation}
  \frac{d\theta_{2}}{dx_{2}} = \frac{d\theta_{1}}{dx_{1}} \theta_{1}^{\lambda-1} \left(\frac{\mu_{1}}{\mu_{2}}\right)^{-\frac{2}{3}},
\end{equation}
where 
\begin{equation}
 \lambda \equiv \left( \frac{n_{1}-3}{n_{2}-3} \right)
\left(\frac{n_{2}+1}{n_{1}+1} \right).
\end{equation}
For comparison with the hydrodynamical simulations we choose the
ratio of mean molecular weight between core and envelope
$\mu_{1} / \mu_{2}=4$.

The combination of polytropic indices in their work is not
suitable for modeling a planet. Here, we have chosen $n_{1}
= 0.01$, $n_{2} = 1$, and $\gamma_{1} = 1+1/n_{1} =
101$, $\gamma_{2} = 1+1/n_{2} = 2$. The reason we did
not use the same polytropic index for the core as in the
simulations is that we want to study the extreme case in which
the entire envelope would be shed, to model the incompressible
core we need a extremely large $\gamma$.

The perturbed polytrope is described by 
\begin{equation}
\omega(x) \equiv P(x)/P_0(x)=(\rho/\rho_0)^\gamma,
\end{equation}
where subscript 0 denotes the unperturbed polytrope. 
The continuity requirements are: 
\begin{equation}
 \omega_{2}=\omega_{1},
\end{equation}
\begin{equation}
 \frac{d \omega_{2}}{d x_{2}}=\frac{d\omega_{1}}{d x_{1}}.
\end{equation}
The overall tendency of the composite polytrope to shrink or
expand is determined by the competition between each component
\citep{Hjellming:1987ys}.

\bibliographystyle{apj.bst}
\bibliography{references}\label{sec:ref}
\end{document}